# Consonant gemination in Italian: the affricate and fricative case

Maria Gabriella Di Benedetto (*), Luca De Nardis (**)

(*) Radcliffe Institute for Advanced Study at Harvard University, Cambridge, USA, and Sapienza University of Rome, Rome, Italy

(**) DIET Department, Sapienza University of Rome
Rome, Italy

{mariagabriella.dibenedetto, luca.denardis}@uniroma1.it

**Abstract**
Consonant gemination in Italian affricates and fricatives was investigated, completing the overall study of gemination of Italian consonants. Results of the analysis of other consonant categories, i.e. stops, nasals, and liquids, showed that closure duration for stops and consonant duration for nasals and liquids, form the most salient acoustic cues to gemination. Frequency and energy domain parameters were not significantly affected by gemination in a systematic way for all consonant classes. Results on fricatives and affricates confirmed the above findings, i.e., that the primary acoustic correlate of gemination is durational in nature and corresponds to a lengthened consonant duration for fricative geminates and a lengthened closure duration for affricate geminates. An inverse correlation between consonant and pre-consonant vowel durations was present for both consonant categories, and also for both singleton and geminate word sets when considered separately. This effect was reinforced for combined sets, confirming the hypothesis that a durational compensation between different phonemes may serve to preserve rhythmical structures. Classification tests of single vs. geminate consonants using the durational acoustic cues as classification parameters confirmed their validity, and highlighted peculiarities of the two consonant classes. In particular, a relatively poor classification performance was observed for affricates, which led to refining the analysis by considering dental vs. non-dental affricates in two different sets. Results support the hypothesis that dental affricates, in Italian, may not appear in intervocalic position as singletons but only in their geminate form.

## 1. Introduction

This work concludes the analysis of gemination in Italian consonants carried out in the framework of the Gemination project GEMMA (Di Benedetto, 2000, GEMMA, 2019) that started at Sapienza in 1992, by addressing the fricatives and affricates consonant classes. Stops were addressed in (Esposito and Di Benedetto, 1999) and liquids and nasals were addressed in the companion paper (Di Benedetto and De Nardis, 2020) that also provides a complete introduction to gemination in Italian.

Gemination in affricates is a particularly challenging topic. Few studies address this consonant class; Abramson (1999) analyzed affricates of Pattani Malay but pointed out that in a pre-test perceptive analysis the percentage of errors was fairly high, and for this reason affricates were eventually discarded.

As regards Italian, the existence of singleton and geminate versions of intervocalic affricates is controversial. A reference study of Italian phonology (Muljacic, 1972) suggests that only non-dental Italian affricates /tʃ/, /dʒ/ may occur in intervocalic position in both singleton and geminate forms, while dental affricates /ts/, /dz/ are always geminated in intervocalic position. Other researchers suggested, however, that intervocalic affricates only exist in their geminated form, and that the distinction between singleton vs. geminate is an artificial construct (Franceschi, 1964). Oppositely, Romeo (1967) suggests that dental affricates may occur as both singleton and geminate, as for example in Gaza (the city) vs. gazza (magpie). This hypothesis is, however, not generally accepted (Muljacic, 1972) on the ground that the two forms are only present in some dialects and cannot be considered as a characteristic feature of Standard Italian. In this investigation, that focuses on Standard Italian, speakers from the area of Rome were selected (see (Di Benedetto and De Nardis, 2020) for details). Interested readers can refer to (Mairano and De Iacovo (2019)) for a comprehensive study of the impact of regional variations of Italian on gemination.

Muljacic view on the non-existence of singleton dental affricates in Standard Italian was later supported by Bertinetto and Loporcaro (2005), reinforcing the consensus on its validity. With the aim of providing additional quantitative evidence on this matter, we decided to include both dental and non-dental affricates in the GEMMA database, with the goal of increasing quantitative evidence some light on the behavior of Italian affricates - with respect to gemination - by analyzing its acoustic properties and testing acoustic cues based on classification tests. The present study includes novel exhaustive statistical analyses of time, frequency, and energy domain parameters for affricates and fricatives, and a thorough comparison of data obtained for affricates and fricatives vs. nasals, liquids, and stops, supported by new statistical tests on stops. In addition, the paper provides new insights on the identification and classification of gemination across different consonant classes.

Section 2 provides a detailed description of the speech materials for affricates. Acoustic measurements and statistical tests are presented in Section 3. Results of acoustic analyses are reported in Section 4. Section 5 presents and discusses results of the classification tests. Finally, Section 6 draws conclusions.

## 2. Speech materials

The speech materials used on this paper belong to the GEMMA database, that includes a complete set of Italian consonants in VCV vs. VCCV words. The database is available under a Creative Commons open source license (GEMMA, 2019); a detailed description of the database is provided in the companion paper (Di Benedetto and De Nardis, 2020).

### 2.1. Affricates and fricatives speech materials

In the Italian language, the set of affricate consonants is /tʃ, dʒ, ts, dz/. As mentioned in the Introduction, the GEMMA database includes words in both forms as shown in Table I, i.e. VCV and VCCV, where the consonant was single /tʃ, dʒ, ts, dz/ or geminate, represented by a double grapheme of the consonant as /tʃtʃ, dʒdʒ, tsts, dzdz/, and the vowel was /i, a, u/. Since both /ts/ and /dz/ are spelled as z in written Italian, the printed cards used during the recording sessions used different representation for the /ts/ and /dz/ consonants, written as TS and DZ, respectively.

Words are symmetrical with respect to vowel. Given the number of speakers (6 speakers), the number of repetitions (3 repetitions), the number of symmetrical vowel contexts (3 vowel contexts), the number of consonants (4 consonants) and the forms (singleton vs. geminate), a total of 6x3x3x4x2=432 words were recorded.

|   | tʃ | | dʒ | | ts | | dz | |
|---|---|---|---|---|---|---|---|---|
| **a** | atʃa | atʃtʃa | adʒa | adʒdʒa | atsa | atstsa | adza | adzdza |
| **i** | itʃi | itʃtʃi | idʒi | idʒdʒi | itsi | itstsi | idzi | idzdzi |
| **u** | utʃu | utʃtʃu | udʒu | udʒdʒu | utsu | utstsu | udzu | udzdzu |

**Table I** - Set of words of the GEMMA database containing affricate consonants. Singleton consonants are indicated by /tʃ, dʒ, ts, dz/. Geminate consonants are indicated by /tʃtʃ, dʒdʒ, tsts, dzdz/.

Regarding fricatives, the set of Italian fricatives is /f, v, s/. These consonants appear in Italian in intervocalic position in both singleton and geminate forms. Table II shows the set of words in the GEMMA database containing fricative consonants, where consonants in the geminated form are in this case as well represented by a double grapheme of the consonant. Given the number of speakers (6 speakers), the number of repetitions (3 repetitions), the number of symmetrical vowel contexts (3 vowel contexts), the number of consonants (3 consonants) and the forms (singleton vs. geminate), a total of 6x3x3x3x2=324 words were recorded.

|   | f | | v | | s | |
|---|---|---|---|---|---|---|
| **a** | afa | affa | ava | avva | asa | assa |
| **i** | ifi | iffi | ivi | ivvi | isi | issi |
| **u** | ufu | uffu | uvu | uvvu | usu | ussu |

**Table II** - Set of words in the GEMMA database containing fricative consonants. Singleton consonants are indicated by /f,v,s/. Geminate consonants are indicated by /ff, vv, ss/.

# 3. Measurements and statistical tests

The analyzed parameters refer to time, frequency, and energy domains. Measurements of the parameters were taken at specific times and frames that are defined in the Section 3.1.1. Time domain parameters are described in Section 3.1.2, while frequency domain and energy domain parameters are described in Sections 3.1.3 and 3.1.4, respectively. The reader is referred to Sections 3.1.1 and 3.1.5 of the companion paper (Di Benedetto and De Nardis, 2020) for a description of the software tools used to collect, process and analyze the data, and of the statistical tests used to determine the statistical significance of the variations of the parameters.

### 3.1.1. Reference times and reference frames

The analysed parameters were measured at specific instants in time, called reference times, that identify abrupt events within the word. The reference times, as shown in Figure 1, are defined as follows:

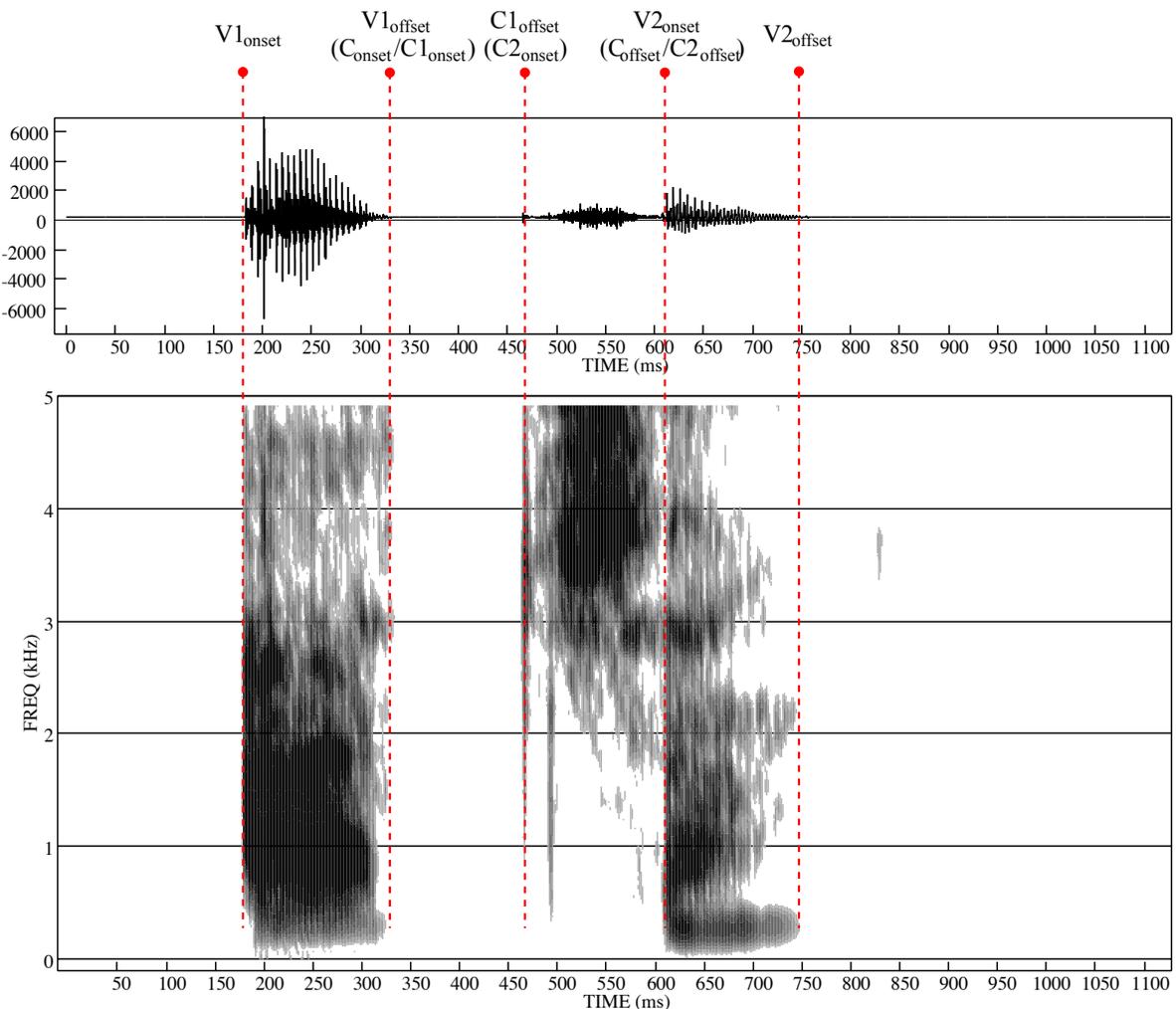

**Figure 1 -** Reference times for the computation of the acoustic parameters: $V1_{onset}$: reference time corresponding to onset of pre-consonant vowel; $V1_{offset}$: offset of pre-consonant vowel, corresponding to onset of the consonant (referred to as $C_{onset}$ for fricatives and $C1_{onset}$ for affricates); $C1_{offset}$: offset of closure for affricates, corresponding to the onset of the fricative part (referred to as $C2_{onset}$); $V2_{onset}$: onset of post-consonant vowel, corresponding to the offset of the consonant (referred to as $C_{offset}$ for fricatives and $C2_{offset}$ for affricates); $V2_{offset}$: offset of post-consonant vowel.

- Vowel 1 onset time ($V1_{onset}$) – The pre-consonant vowel onset time, $V1_{onset}$, was identified by the appearance of a glottal pulse followed by other regular glottal pulses.
- Vowel 1 offset time ($V1_{offset}$) – The pre-consonant vowel offset time, $V1_{offset}$, was identified in a different way for fricatives vs. affricates, given the different articulatory characteristics of those two consonant classes. For fricatives, $V1_{offset}$ was identified as the time at which glottal pulses disappear, corresponding to most energy being concentrated above 1 kHz. For affricates, $V1_{offset}$ was matched with the disappearance of glottal pulses in combination with a sharp decrease in energy due to closure.

- Vowel 2 onset time ($V2_{onset}$) – The post-consonant vowel onset time, $V2_{onset}$, was identified as the reference time at which energy above 1 kHz appears.
- Vowel 2 offset time ($V2_{offset}$) – The post-consonant vowel offset time, $V2_{offset}$, was typically matched with the disappearance of the second and higher formants. In specific cases, mostly with [i] and [u], this reference time was set as the time at which the amplitude of the signal decreased below 90% of its peak value.
- Consonant onset time ($C_{onset}$) – The consonant onset time, $C_{onset}$, is defined for fricative consonants and coincides with $V1_{offset}$.
- Consonant part 1 onset time ($C1_{onset}$) – The presence in the affricate of a closure followed by a frication ([-continuant]) requires splitting the consonant in two parts: C1, corresponding to the closure, and C2, corresponding to the frication. $C1_{onset}$ indicates the onset of C1 and coincides with $V1_{offset}$.
- Consonant part 1 offset ($C1_{offset}$) – The consonant part 1 offset time, $C1_{offset}$, is defined for affricate consonants, and matched to an increased short-term signal energy in combination with the appearance of high frequency components, caused by the release of closure.
- Consonant part 2 onset ($C2_{onset}$) – The consonant part 2 onset time, $C2_{onset}$, labels the onset of the fricative part of affricate consonants, and coincides with $C1_{offset}$.
- Consonant offset ($C_{offset}$) – The consonant offset time $C_{offset}$ is defined for fricative consonants, and coincides with $V2_{onset}$, given the [+continuant] property of fricative consonants.
- Consonant part 2 offset ($C2_{offset}$) – The consonant part 2 offset time, $C2_{offset}$, is defined for affricate consonants, and coincides with $V2_{onset}$.

A set of reference frames, each consisting of 256 samples, was also defined, with respect to reference times. Figure 2 shows the reference frames, that are defined as follows:
- V1 CENTRE – frame located at V1 center, i.e. centered at $\frac{V1_{onset}+V1_{offset}}{2}$;
- V1 OFFSET – frame located at the offset of V1, right before $V1_{offset}$;
- V1-TO-C TRANSITION – frame located at the transition between V1 and C, centered on $V1_{offset}$;
- C ONSET (fricatives only) – frame located at the onset of the consonant, i.e. starting at $V1_{offset}$;
- C1 ONSET (affricates only) – frame located at the onset of closure, i.e. starting at $V1_{offset}$;
- C1 CENTRE (affricates only) – frame located at C1 center, i.e. centered on $\frac{V1_{offset}+C1_{offset}}{2}$;
- C CENTRE (fricatives only) – frame located at C center, i.e. centered on $\frac{V1_{offset}+C_{offset}}{2}$;
- C2 CENTRE (affricates only) – frame located at C2 center, i.e. centered on $\frac{C1_{offset}+C2_{offset}}{2}$;
- C1-TO-C2 TRANSITION (affricates only) – frame located at the C1-C2 transition, centered at $C1_{offset}$;
- C OFFSET (fricatives only) – frame located at the offset of the consonant, i.e. ending at $C_{offset}$;
- C2 OFFSET (affricates only) – frame located at the offset of closure, i.e. ending at $C2_{offset}$;
- V2 ONSET – frame located at the onset of V1, i.e. starting at $V2_{onset}$;
- V2 CENTRE – frame located at the center of V2, i.e. centered at $\frac{V2_{onset}+V2_{offset}}{2}$.

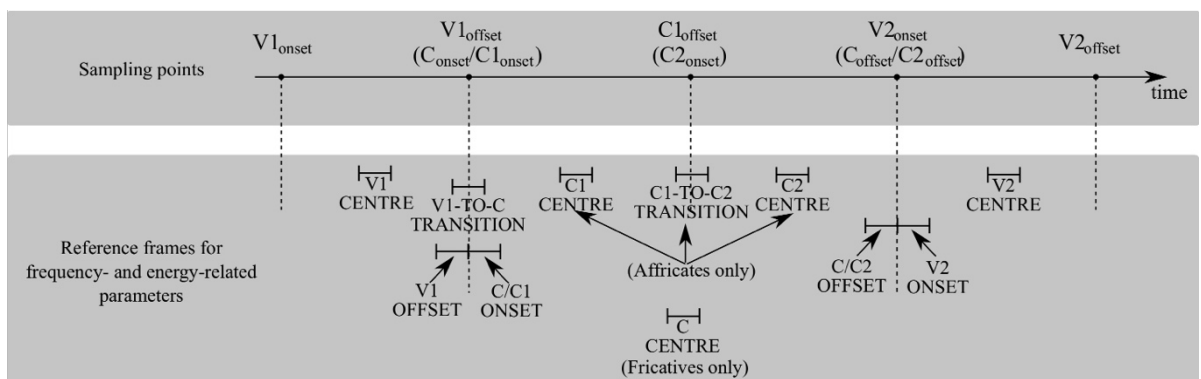

**Figure 2** – Reference frames defined with respect to the reference times of Figure 1. Each reference frame contains 256 samples.

### 3.1.2. Time domain parameters
Figure 3 shows the time domain parameters, defined as follows:
- duration of pre-consonant vowel V1d, defined as $V1d = V1_{offset} - V1_{onset}$;
- duration of closure C1d (for affricates only), defined as $C1d = C1_{offset} - C1_{onset}$;
- duration of frication of consonant C2d (for affricates only), defined as $C2d = C2_{offset} - C2_{onset}$;
- duration of consonant Cd, defined as $Cd = C_{offset} - C_{onset}$; for affricates one has $Cd = C1d + C2d$;
- duration of post consonant vowel V2d, defined as $V2d = V2_{offset} - V2_{onset}$;
- duration of entire word Utd, defined as $Utd = V2_{offset} - V1_{onset}$.

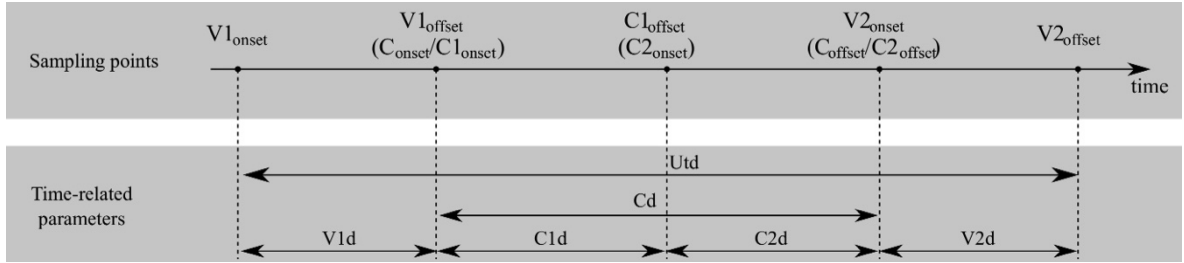

**Figure 3** – Time domain parameters defined with respect to reference times (see Fig. 1). V1d: duration of pre-consonant vowel; C1d: duration of closure (affricates only); C2d: duration of fricative part of consonant (affricates only); Cd: duration of consonant; V2d: duration of post-consonant vowel; Utd: duration of the word.

### 3.1.3. Frequency domain parameters
The following parameters were measured and considered in the analysis:
- Fundamental frequency F0;
- First three formant frequencies F1, F2 and F3.

The above parameters were evaluated with respect to the reference frames as follows (see Figure 2 for reference):
- V1 CENTRE: F0, F1, F2 and F3;
- V1 OFFSET: F0, F1, F2 and F3;
- V1-TO-C TRANSITION: F0, F1, F2 and F3;
- C ONSET: F0 (voiced fricatives only);
- C1 ONSET: F0 (voiced affricates only);
- C1 CENTRE: F0 (voiced affricates only);
- C2 CENTRE: F0 (voiced affricates only);
- C CENTRE: F0 (voiced fricatives only);
- C2 OFFSET: F0 (voiced affricates only);
- C OFFSET: F0 (voiced fricatives only);
- V2 ONSET: F0, F1, F2 and F3;
- V2 CENTRE: F0, F1, F2 and F3.

### 3.1.4. Energy domain parameters
The following energy domain parameters were defined:
- total energy of V1, $E_{totV1}$, defined as $E_{totV1} = \sum |X_i|^2$, where $X_i$ is i-th sample falling in the time interval $[V1_{onset}, V1_{offset}]$;
- average power of V1, defined as $P_{V1} = E_{totV1}/N_{V1}$, where $N_{V1}$ is the number of samples within the interval $[V1_{onset}, V1_{offset}]$;
- total energy of C1, $E_{totC1}$, computed as for V1, but over the interval $[C1_{onset}, C1_{offset}]$, (for affricates only);
- average power of C1, $P_{C1}$, computed from $E_{totC1}$ as for $P_{V1}$, but dividing by the number of samples within the interval $[C1_{onset}, C1_{offset}]$ (for affricates only);
- total energy of C2, $E_{totC2}$, computed as for V1, but over the interval $[C2_{onset}, C2_{offset}]$, (for affricates only);
- average power of C2, indicated as $P_{C2}$ and computed from $E_{totC2}$ as for $P_{V1}$, but dividing by the number of samples within the interval $[C2_{onset}, C2_{offset}]$, (for affricates only);
- total energy of C, $E_{totC}$, computed as for V1, but over the interval $[C_{onset}, C_{offset}]$;
- average power of C, $P_C$, computed from $E_{totC}$ as for $P_{V1}$, but dividing by the number of samples within the interval $[C_{onset}, C_{offset}]$;

- instantaneous energy at V1 CENTRE, indicated as $E_{iV1cent}$, defined as $E_{iV1cent}=\sum|X_i|^2$, where $X_i$ is i-th sample belonging to the V1 CENTRE reference frame;
- instantaneous energy at the transition V1-to-C, $E_{iV1-C}$, computed as for $E_{iV1cent}$ but in the V1-TO-C TRANSITION reference frame;
- instantaneous energy at C CENTRE, $E_{iCcent}$, computed as for $E_{iV1}$ (fricatives only);
- instantaneous energy at C1 CENTRE, $E_{iC1cent}$, computed as for $E_{iV1}$ (affricates only);
- instantaneous energy at C1-TO-C2 TRANSITION, $E_{iC1-C2}$, computed as for $E_{iV1}$ (affricates only);
- instantaneous energy at C2 CENTRE, $E_{iC2cent}$, computed as for $E_{iV1}$ (affricates only);
- instantaneous energy at C CENTRE, $E_{iCcent}$, computed as for $E_{iV1}$ (fricatives only);
- instantaneous energy at C2 OFFSET, $E_{iC2off}$, computed as for $E_{iV1}$ (affricates only);
- instantaneous energy at C OFFSET, $E_{iCoff}$, computed as for $E_{iV1}$ (fricatives only).

All energy domain parameters listed above were expressed in logarithmic form ($10\log_{10}(x)$).

## 4. Results

### 4.1. Results on affricates
#### 4.1.1. Results in the time domain

Figure 4 shows the time domain parameters, V1d, C1d, C2d, Cd, V2d and Utd, averaged over all repetitions and speakers for the affricate consonants [ʧ, ʤ, ts, dz], and the corresponding standard deviations (Table XXVI in Appendix). Values of V1d, C1d and C2d show a general tendency to shorten V1d and lengthen the consonant, both closure section C1d and fricative section C2d, in geminate vs. singleton words. Results also confirm the finding that second vowel duration V2d is not affected by gemination in a systematic form (Di Benedetto and De Nardis, 2020; Esposito and Di Benedetto, 1999). Note that geminate words were slightly longer than singleton ones. The significance of the above trends was investigated by applying a set of statistical tests as described in the following.

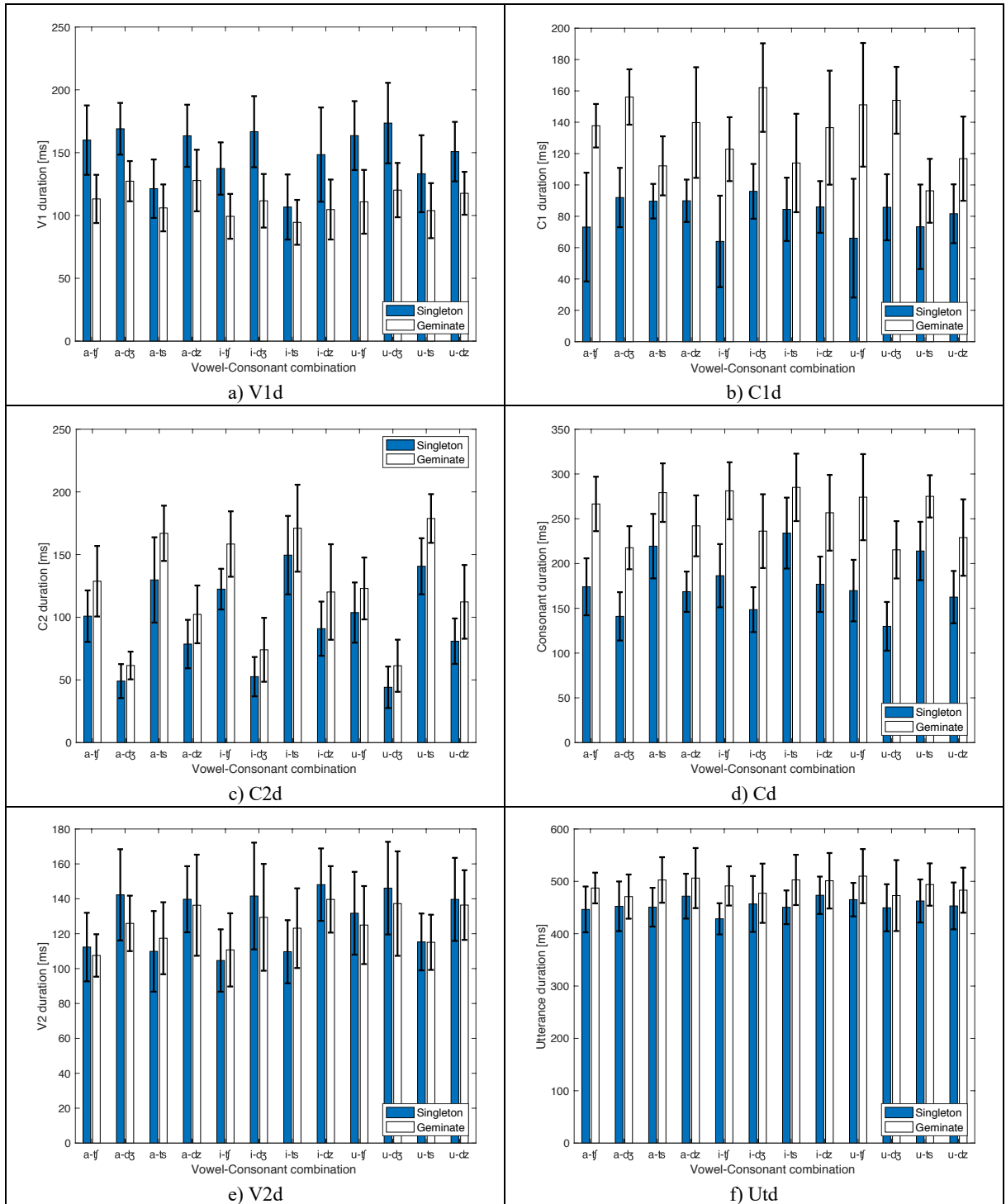

**Figure 4** – Average and standard deviation of time domain parameters for affricate words in singleton vs. geminate forms, averaged over all repetitions and speakers (all values are expressed in milliseconds).

A repeated measurements ANOVA test was performed on female and male speakers data separately, averaged over repetitions. Form (singleton vs. geminate) was used as a between-subjects factor, while Vowel ([a, i, u]) and Consonant ([tʃ, dʒ, ts, dz]) were considered as within-subject factors. Note that the distinction between Form as a between-subjects factor vs. Vowel and Consonant as within-subject factors is not related to the way data were collected, since each speaker recorded all combinations of Form, Vowel and Consonant. The distinction was rather the result of an experiment design choice. Table III shows the test variable F and the corresponding p value for each factor and for the interaction between each within-subjects factor and the between-subjects factor; bold values indicate significant values, with threshold set as $p^*=0.05$.

|  |  | Female | | Male | |
| --- | --- | --- | --- | --- | --- |
|  |  | F | p | F | p |
| V1d | Form | F(1,4)=3.650 | 0.129 | **F(1,4)=8.938** | **0.04** |
|  | Vowel*Form | F(2,8)=0.272 | 0.630 | F(2,8)=1.461 | 0.288 |
|  | Consonant*Form | F(3,12)=1.839 | 0.194 | F(3,12)=3.047 | 0.133 |
|  | Vowel | F(2,8)=1.529 | 0.274 | **F(2,8)=24.932** | **<0.001** |
|  | Consonant | **F(3,12)=10.447** | **0.001** | **F(3,12)=11.249** | **0.001** |
| C1d | Form | **F(1,4)=9.829** | **0.035** | **F(1,4)=129.906** | **<0.001** |
|  | Vowel*Form | F(2,8)=0.706 | 0.471 | F(2,8)=0.236 | 0.795 |
|  | Consonant*Form | F(3,12)=2.265 | 0.133 | F(3,12)=2.992 | 0.073 |
|  | Vowel | F(2,8)=1.098 | 0.379 | F(2,8)=2.918 | 0.112 |
|  | Consonant | **F(3,12)=3.807** | **0.040** | **F(3,12)=3.934** | **0.036** |
| C2d | Form | F(1,4)=6.149 | 0.068 | F(1,4)=4.098 | 0.113 |
|  | Vowel*Form | F(2,8)=0.017 | 0.983 | F(2,8)=0.063 | 0.939 |
|  | Consonant*Form | F(3,12)=1.339 | 0.308 | F(3,12)=0.237 | 0.869 |
|  | Vowel | F(2,8)=1.418 | 0.297 | **F(2,8)=17.778** | **0.001** |
|  | Consonant | **F(3,12)=69.527** | **<0.001** | **F(3,12)=44.101** | **<0.001** |
| V2d | Form | F(1,4)=0.030 | 0.871 | F(1,4)=0.077 | 0.795 |
|  | Vowel*Form | F(2,8)=1.947 | 0.205 | F(2,8)=0.435 | 0.662 |
|  | Consonant*Form | F(3,12)=0.984 | 0.433 | **F(3,12)=4.694** | **0.022** |
|  | Vowel | **F(2,8)=7.896** | **0.013** | F(2,8)=3.081 | 0.102 |
|  | Consonant | **F(3,12)=18.262** | **<0.001** | **F(3,12)=32.903** | **0.005** |
| Utd | Form | F(1,4)=2.362 | 0.199 | F(1,4)=0.938 | 0.388 |
|  | Vowel*Form | F(2,8)=0.372 | 0.701 | F(2,8)=0.163 | 0.853 |
|  | Consonant*Form | F(3,12)=1.783 | 0.204 | F(3,12)=0.448 | 0.723 |
|  | Vowel | F(2,8)=1.741 | 0.236 | F(2,8)=0.996 | 0.411 |
|  | Consonant | F(3,12)=1.687 | 0.222 | F(3,12)=1.058 | 0.403 |
| Cd | Form | **F(1,4)=9.804** | **0.035** | **F(1,4)=63.847** | **0.001** |
|  | Vowel*Form | F(2,8)=0.673 | 0.537 | F(2,8)=0.061 | 0.941 |
|  | Consonant*Form | F(3,12)=1.069 | 0.399 | F(3,12)=2.053 | 0.160 |
|  | Vowel | F(2,8)=3.834 | 0.068 | **F(2,8)=5.018** | **0.039** |
|  | Consonant | **F(3,12)=26.399** | **<0.001** | **F(3,12)=14.723** | **<0.001** |

**Table III** – Results of the repeated measurements multivariate ANOVA test performed on time domain parameters for affricate words. Data were grouped separately for female and male speakers, and averaged over repetitions; test variable F and corresponding probability p at which the null hypothesis can be rejected are presented for the between-subjects factor Form (singleton vs. geminate), for the within-subjects factors Vowel ([a, i, u]) and Consonant ([tʃ, dʒ, ts, dz]), and for the interactions between Form and each within-subject factor; bold characters indicate significantly different values, with threshold set as p*=0.05.

Results in Table III show that gemination has a significant impact on the average value of C1d and Cd for both female and male speakers, and on V1d for male speakers. In the case of C2d, p values close to the selected significance threshold were observed for female speakers, suggesting a weak possible impact of gemination on this parameter as well. No significant variations were observed for V2d and Utd.

Consonant has a very strong impact on the C2d parameter for both female and male speakers; the same behavior can be observed for C1d and V1d. As for the Vowel factor, significant variations can be observed for V2d for female speakers, and for V1d, Cd and C2d for male speakers.

In order to get further insight on the impact of gemination, additional univariate ANOVA tests were carried out for each vowel and consonant separately, considering Form as the only fixed factor. Male and female speakers were in this case combined, since the results presented in Table III highlighted no major differences between the two genders with respect to gemination. Results are shown in Table IV, and show the test variable F and the corresponding probability p of validity of the null hypothesis; values in bold indicate statistically significant variations between singleton vs. geminate groups, with threshold set as p*=0.05.

Results of Table IV confirm that C1d was the parameter most significantly impacted by gemination; the difference of C1d values in singletons vs. geminates groups were in fact significant for all combinations of consonants and vowels. The duration of the pre-consonant vowel V1d also showed significant variations in most cases, although variations were not significant in the case of [ts] combined with [a] and [i]. Note that the variations of V1d with gemination were not significant for female speakers (see Table III); it appears thus that combining female and male speakers data blurred a marked difference between the two groups. A weaker significance was observed for C2d, with significant variations in almost all cases (with the exception of [ts] combined with [i] and [tʃ] combined with [u]). An even weaker impact was observed for Utd, that only showed significant variations consistently across all vowels for consonant [tʃ]. Finally, the second vowel duration V2d did not vary significantly between singletons vs. geminates for any combination of vowels and consonants.

|  |  | a | | | | | i | | | | | u | | | | |
|---|---|---|---|---|---|---|---|---|---|---|---|---|---|---|---|---|
|  |  | V1d | C1d | C2d | V2d | Utd | V1d | C1d | C2d | V2d | Utd | V1d | C1d | C2d | V2d | Utd |
| tʃ | F(1,34) | **34.89** | **53.9** | **11.52** | 0.78 | **10.83** | **34.69** | **49.22** | **24.66** | 0.88 | **31.05** | **35.75** | **43.56** | **5.62** | 0.78 | **9.82** |
|  | p | **1E-06** | **2E-08** | **0.002** | 0.38 | **0.002** | **1E-06** | **4E-08** | **2E-05** | 0.36 | **3E-06** | **9E-07** | **1E-07** | **0.02** | 0.38 | **0.004** |
| dʒ | F(1,34) | **46.04** | **110.5** | **9.15** | **5.14** | 1.54 | **43.36** | **71.55** | **9.23** | 1.43 | 1.25 | **34.23** | **93.40** | **7.55** | 0.88 | 1.48 |
|  | p | **8E-08** | **3E-12** | **0.0047** | **0.03** | 0.22 | **2E-07** | **7E-10** | **0.005** | 0.24 | 0.27 | **1E-06** | **2E-11** | **0.0096** | 0.35 | 0.23 |
| ts | F(1,34) | **4.73** | **19.22** | **15.15** | 1.06 | **14.91** | 2.69 | **11.29** | 3.80 | 3.86 | **14.75** | **10.96** | **8.32** | **29.77** | 0.002 | **5.36** |
|  | p | **0.04** | **1E-04** | **4E-04** | 0.31 | **5E-04** | 0.11 | **0.002** | 0.06 | 0.06 | **5E-04** | **0.002** | **0.007** | **4E-06** | 0.97 | **0.02** |
| dz | F(1,34) | **18.84** | **31.39** | **11.15** | 0.17 | 4.19 | **17.39** | **28.90** | **8.04** | 1.62 | 3.39 | **23.24** | **20.71** | **14.83** | 0.20 | 4.22 |
|  | p | **1E-4** | **3E-06** | **0.002** | 0.68 | 0.05 | **2E-04** | **6E-06** | **0.008** | 0.21 | 0.07 | **3E-05** | **7E-05** | **5E-04** | 0.66 | 0.05 |

**Table IV** - Test variable F and corresponding probability p at which the null hypothesis can be rejected, obtained in the univariate ANOVA test performed on time domain parameters for affricate words using the Form (singleton vs. geminate) as fixed factor, for each combination of consonants [tʃ, dʒ, ts, dz] and vowels [a, i, u]; bold characters indicate significantly different values, with threshold set as p*=0.05.

Next, a Spearman Rank correlation coefficient test was carried out in order to highlight any possible correlation between time domain parameters also related to gemination. Results of the test are presented in Table Va) for singleton and geminated words separately, and in Table Vb) for all combined words.

|  |  | Singleton | | | | | Geminate | | | | |
|---|---|---|---|---|---|---|---|---|---|---|---|
|  |  | V1d s. | C1d s. | C2d s. | V2d s. | Cd s. | V1d g. | C1d g. | C2d g. | V2d g. | Cd g. |
| Singleton | V1d s. | 1.00 | -0.27 | **-0.43** | **0.5** | **-0.57** | | | | | |
|  | C1d s. | -0.27 | 1.00 | -0.20 | 0.03 | 0.35 | | | not significant | | |
|  | C2d s. | **-0.43** | -0.20 | 1.00 | **-0.5** | **0.81** | | | | | |
|  | V2d s. | **0.5** | 0.029 | **-0.5** | 1.00 | **-0.44** | | | | | |
|  | Cd s. | **-0.57** | 0.35 | **0.81** | **-0.44** | 1.00 | | | | | |
| Geminate | V1d g. | | | | | | 1.00 | 0.04 | **-0.32** | **0.55** | **-0.36** |
|  | C1d g. | | | not significant | | | 0.04 | 1.00 | **-0.51** | 0.08 | 0.13 |
|  | C2d g. | | | | | | **-0.32** | **-0.51** | 1.00 | -0.24 | **0.75** |
|  | V2d g. | | | | | | **0.55** | 0.08 | -0.24 | 1.00 | -0.20 |
|  | Cd g. | | | | | | **-0.36** | 0.13 | **0.75** | -0.20 | 1.00 |

|  | V1d | C1d | C2d | V2d | Cd |
|---|---|---|---|---|---|
| V1d | 1.00 | **-0.47** | **-0.47** | **0.47** | **-0.70** |
| C1d | **-0.47** | 1.00 | -0.05 | -0.02 | **0.61** |
| C2d | **-0.47** | -0.05 | 1.00 | **-0.36** | **0.74** |
| V2d | **0.47** | -0.02 | **-0.36** | 1.00 | **-0.28** |
| Cd | **-0.70** | **0.61** | **0.74** | **-0.28** | 1.00 |

a) Separate groups (singleton vs. geminate)      b) Combined

**Table V** - Spearman Rank Correlation Coefficient $r_s$ of time domain parameters for words containing singleton and geminate affricates (Table Va)), and for all words, singleton and geminate combined (Table Vb)). Bold characters indicate significant correlations, with threshold set at p*=0.05.

Note that correlation coefficients close to 0 indicate negligible correlation between parameters, positive coefficients indicate direct correlation, and negative coefficients indicate inverse correlation. From values on Table V one may conclude that a rhythmical compensation effect between C1d, C2d, and Cd on one side, vs. V1d on the other, is present for singleton and combined groups, since $r_s$ is negative for V1d vs. C1d, V1d vs. C2d and V1d vs. Cd. For the group of geminated words, an inverse correlation is observed for V1d vs. C2d and V1d vs. Cd, although weaker than in the other groups, but not for V1d vs. C1d. It can be thus inferred that the rhythmical compensation may not be related to gemination. A test based on the Pearson's correlation coefficient led to similar results, indicating that relationships between parameters, when they do exist, are linear.

### 4.1.2. Results in the frequency domain
Table VI and Table VII show the mean and standard deviation of frequency domain parameters, for female vs. male speakers, singleton vs. geminate forms, and for each vowel, in reference frames: 1) V1 CENTER, 2) V1

OFFSET, 3) V1-TO-C TRANSITION (Table VI) and 4) C1 ONSET, 5) C1 CENTER, 6) C2 CENTER, 7) C2 OFFSET, 8) V2 ONSET, 9) V2 CENTER (Table VII). Values in both tables are averaged over all consonants, speakers and repetitions.

| | | | V1 CENTER | | | | | | | |
| --- | --- | --- | --- | --- | --- | --- | --- | --- | --- | --- |
| | | | Female (Hz) | | | | Male (Hz) | | | |
| | | | F0 | F1 | F2 | F3 | F0 | F1 | F2 | F3 |
| a | Singleton | Mean | 183 | 1068 | 1648 | 2748 | 115 | 849 | 1356 | 2530 |
| | | StD | 39 | 155 | 158 | 327 | 10 | 30 | 41 | 101 |
| | Geminate | Mean | 189 | 1057 | 1626 | 2761 | 124 | 849 | 1349 | 2496 |
| | | StD | 39 | 90 | 170 | 318 | 8 | 38 | 48 | 125 |
| i | Singleton | Mean | 198 | 397 | 2783 | 3555 | 128 | 284 | 2288 | 3261 |
| | | StD | 37 | 73 | 132 | 271 | 13 | 16 | 39 | 141 |
| | Geminate | Mean | 203 | 404 | 2801 | 3577 | 140 | 285 | 2281 | 3275 |
| | | StD | 41 | 80 | 128 | 271 | 11 | 19 | 56 | 156 |
| u | Singleton | Mean | 198 | 394 | 760 | 2837 | 140 | 307 | 650 | 2420 |
| | | StD | 37 | 72 | 55 | 249 | 11 | 25 | 66 | 128 |
| | Geminate | Mean | 207 | 413 | 753 | 2879 | 149 | 302 | 720 | 2391 |
| | | StD | 39 | 74 | 63 | 203 | 9 | 17 | 41 | 140 |
| | | | V1 OFFSET | | | | | | | |
| | | | Female (Hz) | | | | Male (Hz) | | | |
| | | | F0 | F1 | F2 | F3 | F0 | F1 | F2 | F3 |
| a | Singleton | Mean | 177 | 883 | 1734 | 2858 | 111 | 695 | 1449 | 2504 |
| | | StD | 41 | 131 | 130 | 344 | 12 | 102 | 112 | 111 |
| | Geminate | Mean | 155 | 889 | 1633 | 2652 | 123 | 727 | 1449 | 2477 |
| | | StD | 39 | 120 | 165 | 261 | 10 | 91 | 103 | 117 |
| i | Singleton | Mean | 187 | 372 | 2756 | 3479 | 119 | 288 | 2275 | 3222 |
| | | StD | 39 | 73 | 164 | 296 | 12 | 23 | 50 | 185 |
| | Geminate | Mean | 142 | 421 | 2160 | 3085 | 137 | 284 | 2291 | 3256 |
| | | StD | 22 | 184 | 587 | 497 | 12 | 22 | 38 | 201 |
| u | Singleton | Mean | 188 | 372 | 997 | 2819 | 127 | 311 | 891 | 2258 |
| | | StD | 40 | 76 | 58 | 187 | 12 | 35 | 80 | 153 |
| | Geminate | Mean | 155 | 321 | 1397 | 2756 | 143 | 308 | 932 | 2213 |
| | | StD | 20 | 31 | 679 | 423 | 11 | 23 | 68 | 150 |
| | | | V1-TO-C TRANSITION | | | | | | | |
| | | | Female (Hz) | | | | Male (Hz) | | | |
| | | | F0 | F1 | F2 | F3 | F0 | F1 | F2 | F3 |
| a | Singleton | Mean | 174 | 759 | 1764 | 2924 | 111 | 607 | 1495 | 2505 |
| | | StD | 42 | 172 | 123 | 322 | 13 | 108 | 151 | 114 |
| | Geminate | Mean | 183 | 813 | 1791 | 2910 | 121 | 630 | 1494 | 2483 |
| | | StD | 42 | 163 | 142 | 329 | 12 | 83 | 122 | 115 |
| i | Singleton | Mean | 180 | 351 | 2719 | 3445 | 116 | 301 | 2262 | 3146 |
| | | StD | 38 | 72 | 169 | 292 | 12 | 29 | 62 | 220 |
| | Geminate | Mean | 190 | 378 | 2768 | 3443 | 133 | 293 | 2266 | 3214 |
| | | StD | 42 | 82 | 157 | 276 | 13 | 17 | 63 | 184 |
| u | Singleton | Mean | 182 | 363 | 1067 | 2785 | 122 | 309 | 989 | 2217 |
| | | StD | 39 | 78 | 88 | 200 | 15 | 33 | 87 | 173 |
| | Geminate | Mean | 197 | 385 | 1030 | 2814 | 138 | 298 | 1021 | 2154 |
| | | StD | 43 | 83 | 68 | 214 | 13 | 24 | 96 | 145 |

**Table VI** - Mean and Standard Deviation of pitch F0 and formants F1, F2 and F3 in reference frames V1 CENTER, V1 OFFSET and V1-TO-C TRANSITION for affricate words, for female vs. male speakers, averaged over repetitions, speakers and consonants (frequencies are in Hz).

Results indicate an increased F0 average in geminate words for both male and female speakers, in particular in vowels and voiced affricates frames, while no clear effect of gemination was observed on formants.

|   |   |   | C1 ONSET / C1 CENTER / C2 CENTER / C2 OFFSET | | | | | | | |
|---|---|---|---|---|---|---|---|---|---|---|
|   |   |   | Female (Hz) | | | | Male (Hz) | | | |
|   |   |   | F0 | F0 | F0 | F0 | F0 | F0 | F0 | F0 |
| a | Singleton | Mean | 156 | 143 | 134 | 155 | 105 | 101 | 100 | 105 |
|   |   | StD | *29* | *26* | *6* | *25* | *10* | *12* | *17* | *13* |
|   | Geminate | Mean | 158 | 150 | 141 | 163 | 113 | 104 | 102 | 110 |
|   |   | StD | *33* | *23* | *25* | *26* | *12* | *16* | *19* | *16* |
| i | Singleton | Mean | 166 | 148 | 132 | 152 | 113 | 104 | 102 | 105 |
|   |   | StD | *28* | *25* | *2* | *30* | *14* | *17* | *18* | *16* |
|   | Geminate | Mean | 174 | 156 | 138 | 152 | 127 | 108 | 105 | 109 |
|   |   | StD | *30* | *31* | *25* | *30* | *13* | *18* | *19* | *18* |
| u | Singleton | Mean | 172 | 155 | 140 | 144 | 119 | 108 | 103 | 110 |
|   |   | StD | *34* | *26* | *19* | *17* | *16* | *17* | *16* | *19* |
|   | Geminate | Mean | 179 | 148 | 142 | 149 | 129 | 108 | 105 | 109 |
|   |   | StD | *39* | *30* | *29* | *27* | *11* | *12* | *17* | *17* |
|   |   |   | V2 ONSET | | | | | | | |
|   |   |   | Female (Hz) | | | | Male (Hz) | | | |
|   |   |   | F0 | F1 | F2 | F3 | F0 | F1 | F2 | F3 |
| a | Singleton | Mean | 155 | 684 | 1703 | 3007 | 108 | 534 | 1515 | 2415 |
|   |   | StD | *21* | *159* | *134* | *226* | *13* | *44* | *92* | *89* |
|   | Geminate | Mean | 163 | 671 | 1753 | 3041 | 114 | 535 | 1535 | 2452 |
|   |   | StD | *25* | *177* | *124* | *201* | *14* | *51* | *89* | *88* |
| i | Singleton | Mean | 156 | 309 | 2511 | 3152 | 108 | 307 | 2150 | 2959 |
|   |   | StD | *23* | *44* | *218* | *200* | *12* | *15* | *101* | *249* |
|   | Geminate | Mean | 160 | 321 | 2478 | 3140 | 112 | 308 | 2158 | 3004 |
|   |   | StD | *25* | *44* | *197* | *220* | *11* | *19* | *168* | *329* |
| u | Singleton | Mean | 161 | 324 | 1272 | 2866 | 116 | 320 | 1189 | 2194 |
|   |   | StD | *28* | *51* | *268* | *243* | *17* | *21* | *140* | *152* |
|   | Geminate | Mean | 163 | 329 | 1299 | 2819 | 118 | 328 | 1238 | 2175 |
|   |   | StD | *25* | *59* | *260* | *258* | *19* | *28* | *169* | *182* |
|   |   |   | V2 CENTER | | | | | | | |
|   |   |   | Female (Hz) | | | | Male (Hz) | | | |
|   |   |   | F0 | F1 | F2 | F3 | F0 | F1 | F2 | F3 |
| a | Singleton | Mean | 147 | 942 | 1603 | 3019 | 104 | 679 | 1454 | 2414 |
|   |   | StD | *15* | *79* | *122* | *241* | *15* | *65* | *69* | *72* |
|   | Geminate | Mean | 155 | 937 | 1622 | 3050 | 109 | 666 | 1442 | 2426 |
|   |   | StD | *22* | *79* | *131* | *222* | *18* | *62* | *63* | *91* |
| i | Singleton | Mean | 154 | 307 | 2645 | 3197 | 107 | 299 | 2199 | 3057 |
|   |   | StD | *21* | *40* | *170* | *239* | *14* | *19* | *103* | *213* |
|   | Geminate | Mean | 157 | 320 | 2620 | 3208 | 107 | 302 | 2230 | 3081 |
|   |   | StD | *22* | *47* | *153* | *258* | *12* | *15* | *135* | *288* |
| u | Singleton | Mean | 157 | 317 | 890 | 2910 | 113 | 314 | 913 | 2242 |
|   |   | StD | *24* | *52* | *106* | *243* | *18* | *9* | *68* | *167* |
|   | Geminate | Mean | 158 | 338 | 912 | 2830 | 113 | 315 | 927 | 2224 |
|   |   | StD | *22* | *67* | *92* | *180* | *20* | *18* | *66* | *155* |

**Table VII** - Mean and Standard Deviation of pitch F0 and formants F1, F2 and F3 in reference frames V2 ONSET and V2 CENTER, and of pitch F0 in reference frames C1 ONSET, C1 CENTER, C2 CENTER and C2 OFFSET for affricate words, for female vs. male speakers, averaged with respect to repetitions, speakers and consonants (frequencies are in Hz).

A multi-factor univariate ANOVA test using Form, Vowel and Consonant as fixed factors was thus carried out in order to identify significant variations of frequency domain parameters. No significant effect was observed for any of the parameters in consonant frames, i.e. C1 ONSET, C1 CENTER, C2 CENTER and C2 OFFSET. Results obtained for vowel frames, i.e. V1 CENTER, V1 OFFSET, V1-TO-C TRANSITION,V2 ONSET and V2 CENTER, are presented in Table VIII, as a factor vs. parameter matrix; A checked cell indicates a significant difference in the average value of the parameter due to that factor. Results in Table VIII indicate that Form does not cause significant differences of any of the frequency domain parameters for female speakers, while, for male speakers, F0 shows significant differences in the three frames related to the first vowel. In general, Vowel proved to be by far the main factor inducing significant differences in F0 with the expected trend for high vs. low vowels (Ladd, R. and Silverman, K., 1984). Vowel was also, as expected, the only factor inducing significant differences in formants F1, F2 and F3. The factor Consonant led to significant differences only in sporadic cases, in particular in frames V1-TO-C TRANSITION and V2 ONSET, where significant interaction was also present between Vowel and Consonant factors, suggesting that the significant differences due to Consonant might be an artifact of the strong Vowel-Consonant interactions.

Overall, the only effect of gemination on frequency domain parameters seems therefore to be an increase of F0 in V1 for male speakers, but not for female speakers. In general, frequency domain parameters do not seem to provide much information about gemination across speakers of different genders, as also observed for nasals and liquids in (Di Benedetto and De Nardis, 2020).

|  |  | Female |  |  |  | Male |  |  |  |
|---|---|---|---|---|---|---|---|---|---|
|  |  | F0 | F1 | F2 | F3 | F0 | F1 | F2 | F3 |
| **V1 CENTER** | Form |  |  |  |  | X |  |  |  |
|  | Vowel |  | X | X | X |  | X | X | X |
|  | Consonant |  |  |  |  |  |  |  |  |
| **V1 OFFSET** | Form |  |  |  |  | X |  |  |  |
|  | Vowel |  | X | X | X | X | X | X | X |
|  | Consonant |  |  |  |  |  | X |  |  |
| **V1-TO-C TRANSITION** | Form |  |  |  |  | X |  |  |  |
|  | Vowel |  | X | X | X | X | X | X | X |
|  | Consonant |  | X |  |  |  | X |  |  |
| **V2 ONSET** | Form |  |  |  |  |  |  |  |  |
|  | Vowel |  | X | X | X |  | X | X | X |
|  | Consonant |  | X | X |  |  | X | X |  |
| **V2 CENTER** | Form |  |  |  |  |  |  |  |  |
|  | Vowel |  | X | X | X |  | X | X | X |
|  | Consonant |  |  |  |  |  |  |  |  |

**Table VIII** – Results of the multi-factor univariate ANOVA test performed on frequency domain parameters in vowel reference frames V1 CENTER, V1 OFFSET, V1-TO-C TRANSITION, V2 ONSET and V2 CENTER for affricate words using Form, Vowel and Consonant as fixed factors; a checked cell at the intersection between a parameter and a factor indicates a significant difference between average values for the parameter with respect to the factor.

### 4.1.3. Results in the energy domain

Figure 5 and Figure 6 show the average values of energy domain parameters (for a list of parameters and their definitions refer to Section 3.1.4; (the numerical values are presented in Table XXVII in Appendix). Since in the case of energy domain parameters the impact of gender was not expected to be as strong as for frequency domain parameters, results are presented here averaged over all speakers and repetitions.

No clear trend can be observed from the data presented in Figures 5 and 6. A multi-factor univariate ANOVA test was thus performed in order to determine if statistically significative differences between averages exist; test results are presented in Table IX, showing a factor vs. parameter matrix: a checked cell indicates a significant difference in the average value of the parameter due to that factor.

The test considered the fixed factors Form, Vowel, Consonant and Gender, and was executed twice in two different setups. In the first setup, consonants were divided in two groups; voiced affricates [ʤ, ʣ] vs. voiceless affricates [ʧ, ʦ]; this setup was chosen since voiced consonants are typically characterized by higher energy than voiceless consonants. In the second setup, all consonants were merged in one group. Results of the ANOVA tests show that $E_{totC}$ shows significant variations for voiced, voiceless and combined consonants, in agreement with results for nasals and liquids presented in (Di Benedetto and De Nardis, 2020). No other energy-related parameter presents significant variations due to gemination in all groups, although other parameters do so in only some of the groups: these are $E_{totC2}$ (voiced and voiceless), $E_{iC1cent}$ (voiced and combined), $P_{V1}$, $P_{C2}$, $E_{iV1cent}$ and $E_{iC2off}$ (all for voiceless and combined groups).

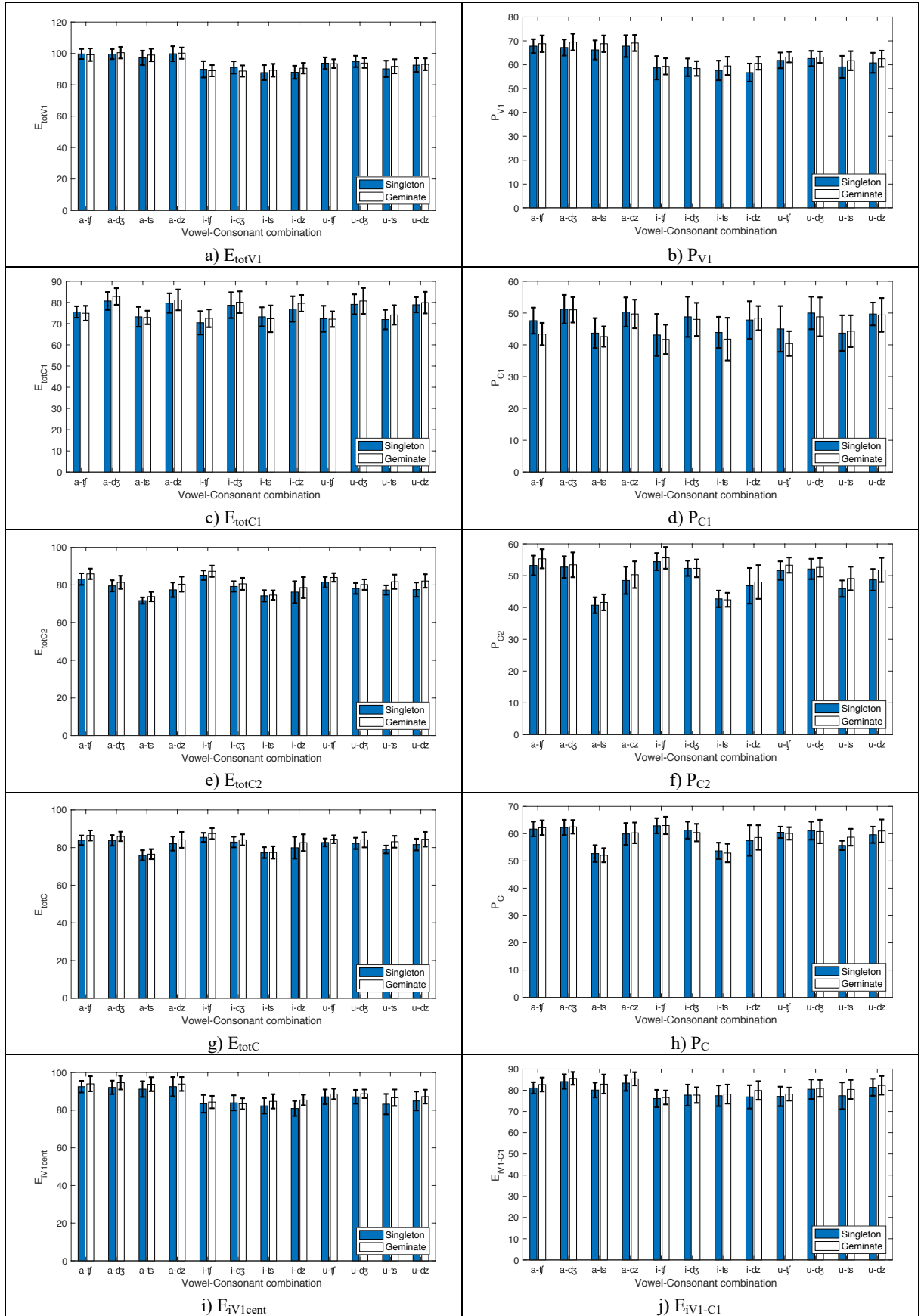

**Figure 5** – Average and standard deviation of energy domain parameters $E_{totV1}$, $P_{V1}$, $E_{totC1}$, $P_{C1}$, $E_{totC2}$, $P_{C2}$, $E_{totC}$, $P_C$, $E_{iV1cent}$ and $E_{iV1-C1}$ for each combination of consonants [tʃ, dʒ, ts, dz], vowels [a, i, u] and singleton vs. geminate form, averaged over repetitions and speakers (values are in logarithmic form)

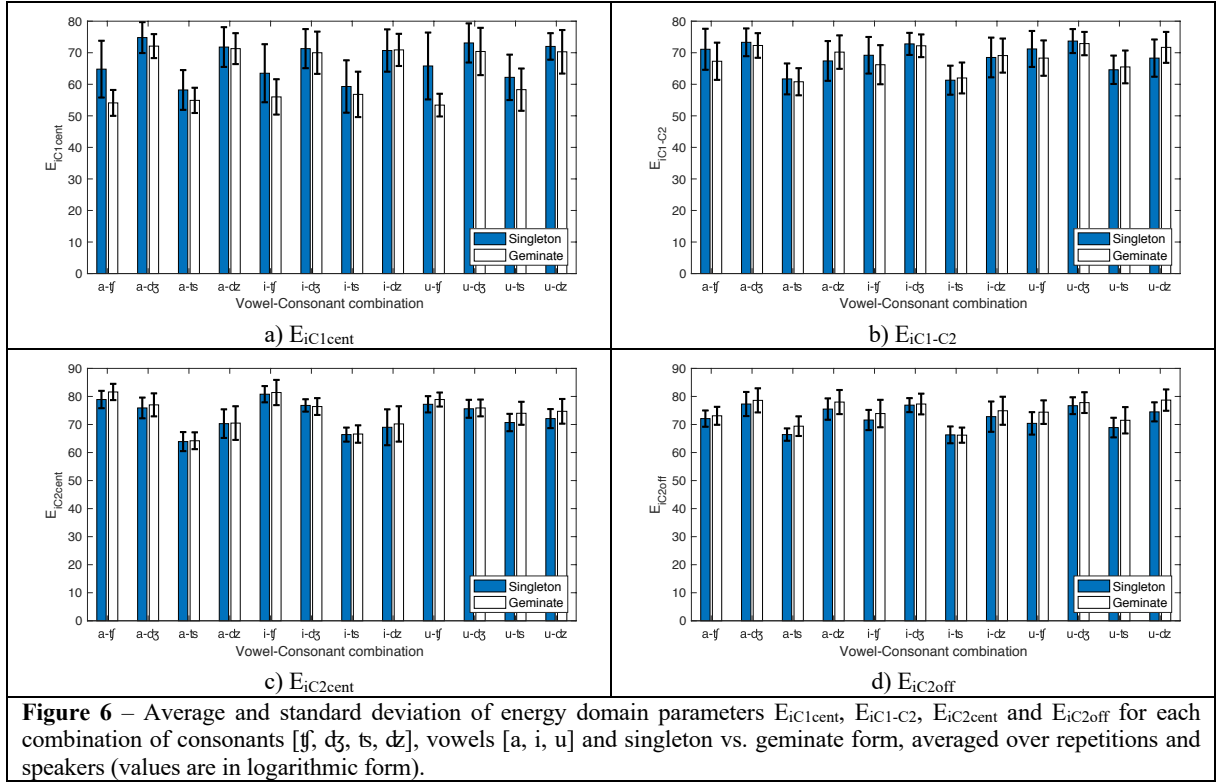

**Figure 6** – Average and standard deviation of energy domain parameters $E_{iC1cent}$, $E_{iC1-C2}$, $E_{iC2cent}$ and $E_{iC2off}$ for each combination of consonants [tʃ, dʒ, ts, dz], vowels [a, i, u] and singleton vs. geminate form, averaged over repetitions and speakers (values are in logarithmic form).

As for the other fixed factors, Vowel led to significant differences in all tests for parameters measured on the first vowel and on the closure ($E_{totV1}$, $P_{V1}$, $E_{iV1cent}$ and $E_{iV1-C1}$) while the Consonant factor led to significant differences for all parameters measured on C1, C2 and C, except for $P_C$ of voiced consonants. Finally, the Gender factor led to significant variations consistent across all three cases for parameters related to V1.

|  |  | $E_{totV1}$ | $P_{V1}$ | $E_{totC1}$ | $P_{C1}$ | $E_{totC2}$ | $P_{C2}$ | $E_{totC}$ | $P_C$ | $E_{iV1cent}$ | $E_{iV1-C1}$ | $E_{iC1cent}$ | $E_{iC1-C2}$ | $E_{iC2cent}$ | $E_{iC2off}$ |
|---|---|---|---|---|---|---|---|---|---|---|---|---|---|---|---|
| **Voiced** | **Form** |  |  |  |  | X |  | X |  |  |  |  | X |  |  |
|  | **Vowel** | X | X |  |  | X |  |  |  | X | X |  |  |  |  |
|  | **Cons.** |  |  | X | X | X | X | X |  |  |  | X | X | X | X |
|  | **Gender** | X | X | X | X |  |  |  |  | X | X |  | X |  |  |
| **Voiceless** | **Form** |  | X |  |  | X | X | X |  | X |  |  |  |  | X |
|  | **Vowel** | X | X |  |  | X | X | X | X | X | X |  |  | X | X |
|  | **Cons.** |  |  | X | X | X | X | X |  |  | X | X | X | X | X |
|  | **Gender** | X | X | X | X | X |  |  |  | X |  | X | X | X |  |
| **Voiced/ voiceless combined** | **Form** |  | X |  |  | X | X | X |  | X |  | X |  |  | X |
|  | **Vowel** | X | X |  |  |  |  |  |  | X | X |  |  | X | X |
|  | **Cons.** |  |  | X | X | X | X | X | X |  | X | X | X | X | X |
|  | **Gender** | X | X | X | X |  |  |  |  | X |  | X |  |  |  |

**Table IX** - Results of the multi-factor univariate ANOVA test performed on energy domain parameters using Form, Vowel, Consonant and Gender for voiced affricates [dʒ, dz], for voiceless affricates [tʃ, ts] and for all combined affricate words; a checked cell indicates a significant difference between average values for the parameter with respect to the factor.

## 4.2. Results on fricatives

### 4.2.1. Results in the time domain

The acoustic time domain parameters listed in Section 3.1.2 were computed for each of the 162 singleton and 162 geminate fricative words. Results are presented in Figure 7, that shows the average values and standard deviations of V1d, Cd, V2d and Utd for all combinations of vowels [a, i, u] and consonants [f, v, s] in geminate vs. singleton forms, averaged over all repetitions and speakers (the numerical values are presented in Table XXVIII in Appendix). Figure 7 shows that generally speaking fricatives behave like affricates regarding V1d and Cd; V1d tends to decrease with gemination, while the opposite happens to Cd. No clear trend can be observed for V2d and Utd.

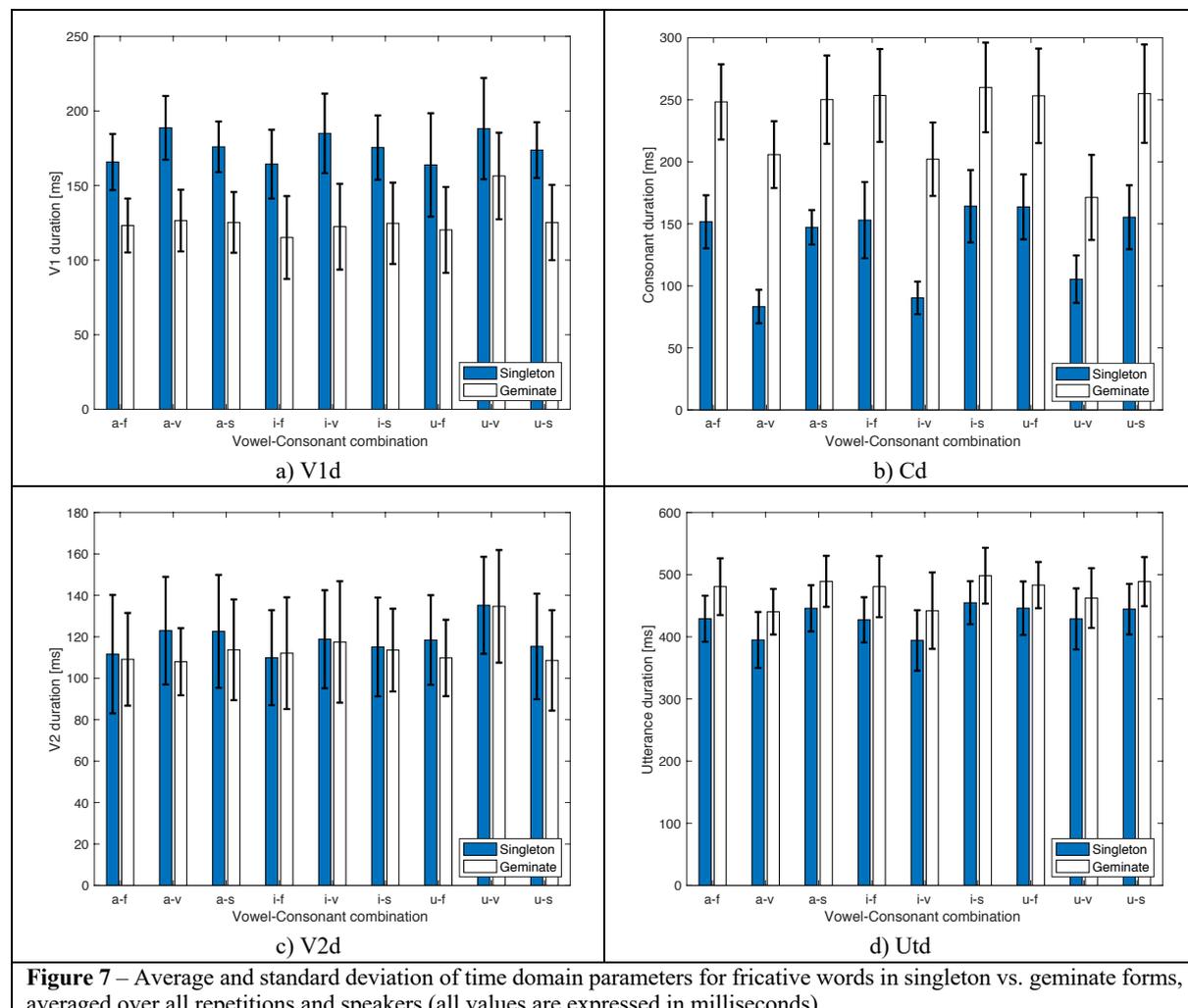

**Figure 7** – Average and standard deviation of time domain parameters for fricative words in singleton vs. geminate forms, averaged over all repetitions and speakers (all values are expressed in milliseconds).

Following a similar approach as in Section 4.1.1, a repeated measurements ANOVA test was performed on female and male speakers data separately, after averaging over repetitions, using Form (singleton vs. geminate) as a between-subjects factor, and Vowel ([a, i, u]) and Consonant ([f, v, s]) as within-subjects factors. Results of the test are presented in Table X. For each parameter, Table X shows the test variable F and the corresponding p value for each factor and for the interaction between each within-subjects factor and the between-subjects factor. Bold values indicate significant variations, with threshold set at $p^*=0.05$.

In terms of gemination, results in Table X highlight a significant variation of Cd for both female and male speakers, while only male speakers show a significant variation of V1d. No significant variations were observed for V2d and for Utd.

As for other factors, Consonant has a strong impact on Cd for both female and male speakers, as well as on V1d and Utd. For V2d, a significant difference was observed for males, but not for females. No significant interaction was observed between Consonant and Form, suggesting that, in fricatives, these are mutually independent. Finally, Vowel was significant only for Cd of male speakers, although a significant interaction between Vowel and Form was also observed, and therefore the impact of Vowel on Cd may be an artifact.

Following the same approach adopted for affricates in Section 4.1.1 and for nasals and liquids in (Di Benedetto and De Nardis, 2020) additional univariate ANOVA tests for the Form factor (gemination) were carried out for each combination of vowel and consonant separately, on combined female and male speakers data. Results, presented in Table XI, confirm Table X. Consonant duration Cd is the parameter showing the most significant variations, followed by V1d; the variation caused by gemination is significant for both parameters for all combinations of vowels and consonants. A weak significance also appears for Utd, except for words including [v]. Finally, no significant variation was observed for V2d.

|     |                | Female |       | Male  |       |
|-----|----------------|--------|-------|-------|-------|
|     |                | F      | p     | F     | p     |
| V1d | Form           | F(1,4)=4.580 | 0.099 | **F(1,4)=17.783** | **0.014** |
|     | Vowel*Form     | F(2,8)=0.586 | 0.579 | F(2,8)=1.216 | 0.346 |
|     | Consonant*Form | F(2,8)=1.038 | 0.397 | F(2,8)=0.251 | 0.784 |
|     | Vowel          | F(2,8)=0.496 | 0.627 | F(2,8)=1.618 | 0.257 |
|     | Consonant      | **F(2,8)=7.235** | **0.016** | **F(2,8)=5.747** | **0.028** |
| Cd  | Form           | **F(1,4)=11.769** | **0.027** | **F(1,4)=48.642** | **0.002** |
|     | Vowel*Form     | **F(2,8)=9.066** | **0.009** | **F(2,8)=13.055** | **0.003** |
|     | Consonant*Form | F(2,8)=0.404 | 0.681 | F(2,8)=2.321 | 0.160 |
|     | Vowel          | F(2,8)=3.079 | 0.102 | **F(2,8)=4.574** | **0.047** |
|     | Consonant      | **F(2,8)=61.421** | **<0.001** | **F(2,8)=196.658** | **<0.001** |
| V2d | Form           | F(1,4)=0.044 | 0.845 | F(1,4)=0.105 | 0.762 |
|     | Vowel*Form     | F(2,8)=1.005 | 0.408 | F(2,8)=3.995 | 0.063 |
|     | Consonant*Form | F(2,8)=0.454 | 0.651 | F(2,8)=1.366 | 0.309 |
|     | Vowel          | F(2,8)=3.115 | 0.100 | F(2,8)=3.373 | 0.087 |
|     | Consonant      | F(2,8)=1.363 | 0.310 | **F(2,8)=11.318** | **0.005** |
| Utd | Form           | F(1,4)=5.209 | 0.085 | F(1,4)=0.991 | 0.376 |
|     | Vowel*Form     | F(2,8)=1.349 | 0.313 | F(2,8)=0.701 | 0.524 |
|     | Consonant*Form | F(2,8)=0.688 | 0.524 | F(2,8)=0.329 | 0.729 |
|     | Vowel          | F(2,8)=3.167 | 0.097 | F(2,8)=1.815 | 0.224 |
|     | Consonant      | **F(2,8)=30.158** | **<0.001** | **F(2,8)=8.217** | **0.011** |

**Table X** – Results of the repeated measurements ANOVA test performed on time domain parameters, separately on female and male speakers data, averaged over repetitions. Test variable F and corresponding probability p at which the null hypothesis can be rejected are presented for the between-subjects factor Form (singleton vs. geminate), for the within-subjects factors Vowel ([a, i, u]) and Consonant ([f, v, s]), and for their interactions. Bold characters indicate significant variations, with threshold set at p*=0.05.

|   |        | a |   |   |   | i |   |   |   | u |   |   |   |
|---|--------|-----|------|------|------|------|------|------|------|------|------|------|------|
|   |        | V1d | Cd   | V2d  | Utd  | V1d  | Cd   | V2d  | Utd  | V1d  | Cd   | V2d  | Utd  |
| f | F(1,34) | **47.85** | **122.16** | 0.08 | **13.9** | **33.43** | **77.59** | 0.07 | **13.74** | **16.85** | **67.53** | 1.67 | **7.73** |
|   | p      | **6E-08** | **9E-13** | 0.77 | **7E-04** | **2E-06** | **3E-10** | 0.8 | **7E-04** | **2E-04** | **1E-09** | 0.2050 | **0.009** |
| v | F(1,34) | **78.6** | **297.2** | **4.32** | **10.99** | **45.82** | **214.21** | 0.024 | **6.72** | **9.1** | **50.71** | 0.0035 | 4.30 |
|   | p      | **2E-10** | **2E-18** | **0.05** | **0.0022** | **9E-08** | **3E-16** | 0.88 | **0.01** | **0.005** | **3E-08** | 0.95 | 0.05 |
| s | F(1,34) | **65.65** | **130.7** | 1.06 | **11.07** | **38.38** | **76.51** | 0.04 | **10.59** | **43.01** | **79.93** | 0.66 | **11.01** |
|   | p      | **2E-09** | **3E-13** | 0.31 | **0.0021** | **5E-07** | **3E-10** | 0.84 | **0.003** | **2E-07** | **2E-10** | 0.42 | **0.002** |

**Table XI** – Test variable F and corresponding probability p at which the null hypothesis can be rejected obtained in the univariate ANOVA test performed on time domain parameters for words containing fricatives using Form (singleton vs. geminate) as fixed factor, for each combination of consonants [f, v, s] and vowels [a, i, u]. Bold characters indicate significantly different values, with threshold set at p*=0.05.

Next, the Spearman Rank Correlation Coefficient $r_s$ for both singleton and geminate groups was evaluated, by first considering singleton and geminate sets separately, and then combined, with results presented in Table XIIa) and Table XIIb), respectively.

Table XIIa) shows that within each group, both singletons and geminates, an increased consonant duration is associated with a shorter V1 and V2, and vice versa, suggesting that this effect is present irrespective of gemination. Results in Table XIIb) on combined words show an even stronger negative correlation between V1d and Cd, in analogy with affricates (Table V). As in the case of affricates, a Pearson's correlation test led to similar values for correlation coefficients.

|  | | Singleton | | | Geminate | | |
|---|---|---|---|---|---|---|---|
|  | | V1d s. | Cd s. | V2d s. | V1d g. | Cd g. | V2d g. |
| Singleton | V1d s. | 1.00 | -0.38 | 0.53 | | | |
| | Cd s. | -0.38 | 1.00 | -0.26 | colspan not significant | | |
| | V2d s. | 0.53 | -0.26 | 1.00 | | | |
| Geminate | V1d g. | | | | 1.00 | -0.46 | 0.65 |
| | Cd g. | colspan not significant | | | -0.46 | 1.00 | -0.26 |
| | V2d g. | | | | 0.65 | -0.26 | 1.00 |

|  | V1d | Cd | V2d |
|---|---|---|---|
| V1d | 1.00 | -0.76 | 0.48 |
| Cd | -0.76 | 1.00 | -0.24 |
| V2d | 0.48 | -0.24 | 1.00 |

a) Separate groups (singleton vs. geminate)    b) Combined

**Table XII** – Spearman Rank Correlation Coefficient $r_s$ of time domain parameters for singleton and geminate fricative words separately (Table XIIa)) and on all words, singleton and geminate combined (Table XIIb)). Bold characters indicate significant correlations, with threshold set at p*=0.05.

### 4.2.2. Results in the frequency domain

Average value and standard deviation of frequency domain parameters F0, F1, F2 and F3 measured in reference frames related to the first vowel (V1 CENTER, V1 OFFSET, and V1-TO-C TRANSITION) are shown in Table XIII, while Table XIV shows average values and standard deviations of F0 in reference frames related to the consonant (C ONSET, C CENTER, and C OFFSET) and of F0, F1, F2 and F3 to the second vowel (V2 ONSET and V2 CENTER). Data in both tables were obtained for female vs. male speakers separately, and for each combination of vowels [a, i, u] and forms (singleton vs. geminate), averaged over all speakers, consonants and repetitions.

| | | | colspan="8" V1 CENTER | | | | | | | |
|---|---|---|---|---|---|---|---|---|---|---|
| | | | colspan="4" Female (Hz) | | | | colspan="4" Male (Hz) | | | |
| | | | F0 | F1 | F2 | F3 | F0 | F1 | F2 | F3 |
| a | Singleton | Mean | 178 | 1054 | 1522 | 2753 | 114 | 824 | 1306 | 2602 |
| | | StD | 34 | 113 | 167 | 343 | 14 | 28 | 42 | 119 |
| | Geminate | Mean | 186 | 1060 | 1591 | 2718 | 117 | 823 | 1270 | 2594 |
| | | StD | 36 | 99 | 114 | 283 | 12 | 26 | 44 | 152 |
| i | Singleton | Mean | 195 | 380 | 2796 | 3569 | 128 | 289 | 2281 | 3274 |
| | | StD | 42 | 71 | 153 | 285 | 13 | 19 | 49 | 149 |
| | Geminate | Mean | 199 | 403 | 2771 | 3515 | 135 | 285 | 2232 | 3209 |
| | | StD | 37 | 71 | 121 | 303 | 13 | 21 | 82 | 134 |
| u | Singleton | Mean | 199 | 394 | 724 | 2692 | 140 | 309 | 648 | 2408 |
| | | StD | 42 | 77 | 63 | 360 | 12 | 13 | 55 | 138 |
| | Geminate | Mean | 211 | 413 | 759 | 2799 | 143 | 316 | 665 | 2364 |
| | | StD | 47 | 86 | 64 | 265 | 12 | 23 | 46 | 99 |
| | | | colspan="8" V1 OFFSET | | | | | | | |
| | | | colspan="4" Female (Hz) | | | | colspan="4" Male (Hz) | | | |
| | | | F0 | F1 | F2 | F3 | F0 | F1 | F2 | F3 |
| a | Singleton | Mean | 154 | 918 | 1536 | 2796 | 110 | 714 | 1200 | 2511 |
| | | StD | 63 | 96 | 195 | 402 | 17 | 99 | 103 | 79 |
| | Geminate | Mean | 181 | 946 | 1511 | 2741 | 114 | 743 | 1183 | 2492 |
| | | StD | 41 | 63 | 181 | 386 | 15 | 38 | 97 | 113 |
| i | Singleton | Mean | 179 | 350 | 2679 | 3327 | 117 | 293 | 2284 | 3126 |
| | | StD | 41 | 78 | 140 | 214 | 15 | 29 | 58 | 163 |
| | Geminate | Mean | 190 | 376 | 2726 | 3367 | 130 | 301 | 2253 | 3170 |
| | | StD | 43 | 78 | 188 | 338 | 15 | 27 | 80 | 101 |
| u | Singleton | Mean | 184 | 353 | 846 | 2215 | 121 | 330 | 778 | 2338 |
| | | StD | 43 | 87 | 155 | 690 | 15 | 34 | 190 | 62 |
| | Geminate | Mean | 196 | 394 | 852 | 2454 | 131 | 335 | 749 | 2364 |
| | | StD | 50 | 91 | 130 | 360 | 16 | 45 | 180 | 79 |

|   |   |   | \multicolumn{8}{c}{V1-TO-C TRANSITION} |
|---|---|---|---|---|---|---|---|---|---|---|
|   |   |   | \multicolumn{4}{c}{Female (Hz)} | \multicolumn{4}{c}{Male (Hz)} |
|   |   |   | F0 | F1 | F2 | F3 | F0 | F1 | F2 | F3 |
| a | Singleton | Mean | 165 | 840 | 1559 | 2785 | 107 | 677 | 1205 | 2528 |
|   |           | StD  | 34  | 103 | 224  | 385  | 18  | 118 | 165  | 106  |
|   | Geminate  | Mean | 177 | 878 | 1587 | 2758 | 112 | 655 | 1176 | 2500 |
|   |           | StD  | 42  | 114 | 223  | 351  | 17  | 85  | 188  | 172  |
| i | Singleton | Mean | 174 | 337 | 2592 | 3220 | 113 | 298 | 2283 | 3063 |
|   |           | StD  | 38  | 78  | 200  | 244  | 15  | 27  | 62   | 200  |
|   | Geminate  | Mean | 183 | 364 | 2578 | 3266 | 124 | 300 | 2249 | 3143 |
|   |           | StD  | 45  | 97  | 216  | 336  | 17  | 26  | 97   | 244  |
| u | Singleton | Mean | 177 | 342 | 873  | 1905 | 106 | 311 | 789  | 2190 |
|   |           | StD  | 42  | 86  | 164  | 436  | 40  | 26  | 231  | 284  |
|   | Geminate  | Mean | 188 | 390 | 887  | 2463 | 127 | 341 | 807  | 2360 |
|   |           | StD  | 45  | 100 | 148  | 528  | 48  | 56  | 203  | 89   |

**Table XIII** – Mean value and standard deviation of F0 and formants F1, F2 and F3 in reference frames V1 CENTER, V1 OFFSET and V1-TO-C TRANSITION for fricative words for female vs. male speakers, averaged over repetitions, speakers and consonants (values are in Hz).

A multi-factor univariate ANOVA test was performed on frequency domain parameters using Form, Vowel and Consonant as fixed factors; results for reference frames related to vowels (V1 CENTER, V1 OFFSET, V1-TO-C TRANSITION, V2 ONSET and V2 CENTER) are presented in Table XV, where a checked cell indicates a significant difference between average values for the parameter with respect to the factor. F0 did not show significant variations for any combination of factors (Form, Vowel, Consonant) and groups of speakers (Male and Female), in consonant frames C ONSET, C CENTER and C OFFSET, and the corresponding table is thus omitted. Results in Table XV show that gemination does not lead to statistically significant variations for any frequency domain parameter. Vowel was the only factor leading to significant differences of F1, F2 and F3 for both female and male speakers and, to a much lower extent, of F0 (only in the V1 CENTER frame for male speakers). Consonant led sporadically to significant differences in F2 but, in all instances, this corresponded to a significant interaction between Vowel and Consonant factors, suggesting that the significance of the Consonant factor for F2 could be an artifact caused by such interaction.

|   |   |   | \multicolumn{6}{c}{C ONSET / C CENTER / C OFFSET} |
|---|---|---|---|---|---|---|---|---|
|   |   |   | \multicolumn{3}{c}{Female (Hz)} | \multicolumn{3}{c}{Male (Hz)} |
|   |   |   | F0 | F0 | F0 | F0 | F0 | F0 |
| a | Singleton | Mean | 154 | 151 | 148 | 104 | 101 | 104 |
|   |           | StD  | 36  | 40  | 24  | 18  | 18  | 18  |
|   | Geminate  | Mean | 162 | 141 | 137 | 106 | 101 | 105 |
|   |           | StD  | 40  | 26  | 13  | 13  | 18  | 24  |
| i | Singleton | Mean | 161 | 147 | 150 | 108 | 104 | 105 |
|   |           | StD  | 31  | 36  | 32  | 13  | 14  | 14  |
|   | Geminate  | Mean | 173 | 148 | 154 | 123 | 104 | 108 |
|   |           | StD  | 45  | 36  | 36  | 17  | 24  | 24  |
| u | Singleton | Mean | 161 | 152 | 151 | 117 | 111 | 112 |
|   |           | StD  | 39  | 40  | 29  | 24  | 23  | 23  |
|   | Geminate  | Mean | 176 | 148 | 147 | 120 | 109 | 111 |
|   |           | StD  | 39  | 27  | 18  | 23  | 23  | 23  |

|   |   |   | \multicolumn{8}{c}{V2 ONSET} |
|---|---|---|---|---|---|---|---|---|---|---|
|   |   |   | \multicolumn{4}{c}{Female (Hz)} | \multicolumn{4}{c}{Male (Hz)} |
|   |   |   | F0 | F1 | F2 | F3 | F0 | F1 | F2 | F3 |
| a | Singleton | Mean | 155 | 788 | 1459 | 2797 | 112 | 582 | 1192 | 2459 |
|   |           | StD  | 19  | 83  | 144  | 357  | 19  | 39  | 145  | 123  |
|   | Geminate  | Mean | 161 | 728 | 1497 | 2843 | 112 | 555 | 1168 | 2427 |
|   |           | StD  | 23  | 72  | 146  | 376  | 16  | 35  | 140  | 127  |
| i | Singleton | Mean | 155 | 306 | 2626 | 3242 | 114 | 298 | 2233 | 3049 |
|   |           | StD  | 17  | 29  | 122  | 220  | 18  | 18  | 69   | 182  |
|   | Geminate  | Mean | 163 | 339 | 2560 | 3161 | 116 | 303 | 2158 | 2918 |
|   |           | StD  | 25  | 42  | 168  | 334  | 19  | 13  | 102  | 319  |

|   |   |      |     |     | V2 CENTER |      |     |     |      |      |
|---|---|------|-----|-----|-----------|------|-----|-----|------|------|
|   |   |      |     | Female (Hz) |     |      |     | Male (Hz) |      |      |
|   |   |      | F0  | F1  | F2        | F3   | F0  | F1  | F2   | F3   |
| u | Singleton | Mean | 153 | 347 | 843 | 2381 | 116 | 326 | 810 | 2258 |
|   |           | StD  | 15  | 36  | 99  | 460  | 14  | 21  | 217 | 164  |
|   | Geminate  | Mean | 160 | 355 | 910 | 2437 | 117 | 324 | 830 | 2354 |
|   |           | StD  | 20  | 47  | 206 | 397  | 17  | 25  | 238 | 235  |
| a | Singleton | Mean | 145 | 950 | 1453 | 2900 | 104 | 715 | 1290 | 2496 |
|   |           | StD  | 15  | 80  | 90   | 301  | 20  | 51  | 114  | 148  |
|   | Geminate  | Mean | 150 | 957 | 1474 | 2900 | 106 | 680 | 1255 | 2444 |
|   |           | StD  | 17  | 81  | 83   | 357  | 18  | 65  | 105  | 108  |
| i | Singleton | Mean | 150 | 319 | 2788 | 3464 | 110 | 299 | 2245 | 3100 |
|   |           | StD  | 17  | 43  | 218  | 255  | 19  | 20  | 59   | 141  |
|   | Geminate  | Mean | 161 | 329 | 2708 | 3289 | 112 | 300 | 2207 | 2986 |
|   |           | StD  | 23  | 35  | 326  | 382  | 21  | 21  | 132  | 247  |
| u | Singleton | Mean | 157 | 336 | 787  | 2661 | 110 | 312 | 678  | 2350 |
|   |           | StD  | 22  | 68  | 79   | 321  | 17  | 24  | 112  | 251  |
|   | Geminate  | Mean | 160 | 342 | 795  | 2665 | 106 | 325 | 732  | 2343 |
|   |           | StD  | 21  | 55  | 85   | 440  | 40  | 33  | 134  | 225  |

**Table XIV** - Mean average and standard deviation of F0, formants F1, F2 and F3 in reference frames V2 ONSET and V2 CENTER, and of F0 in reference frames C ONSET, C CENTER and C OFFSET for fricative words for female vs. male speakers, averaged with respect to repetitions, speakers and consonants (values are in Hz).

|   |   | Female |   |   |   | Male |   |   |   |
|---|---|--------|---|---|---|------|---|---|---|
|   |   | F0 | F1 | F2 | F3 | F0 | F1 | F2 | F3 |
| **V1 CENTER** | Form |   |   |   |   |   |   |   |   |
|   | Vowel |   | X | X | X | X | X | X | X |
|   | Consonant |   |   |   |   |   |   |   |   |
| **V1 OFFSET** | Form |   |   |   |   |   |   |   |   |
|   | Vowel |   | X | X | X |   | X | X | X |
|   | Consonant |   |   | X |   |   |   | X |   |
| **V1-TO-C TRANSITION** | Form |   |   |   |   |   |   |   |   |
|   | Vowel |   | X | X | X |   | X | X | X |
|   | Consonant |   |   | X |   |   | X | X | X |
| **V2 ONSET** | Form |   |   |   |   |   |   |   |   |
|   | Vowel |   | X | X | X |   | X | X | X |
|   | Consonant |   | X | X |   |   |   | X |   |
| **V2 CENTER** | Form |   |   |   |   |   |   |   |   |
|   | Vowel |   | X | X | X |   | X | X | X |
|   | Consonant |   |   |   |   |   |   | X |   |

**Table XV** – Results of the multi-factor univariate ANOVA test performed on frequency domain parameters in reference frames related to vowels (V1 CENTER, V1 OFFSET, V1-TO-C TRANSITION, V2 ONSET and V2 CENTER) for fricative words using Form, Vowel and Consonant as fixed factors; a checked cell indicates a significant difference between average values for the parameter with respect to the factor. Results for consonant frames C ONSET, C CENTER and C OFFSET are not presented since no significant variation was detected.

#### 4.2.3. Results in the energy domain

Figure 8 shows average value and standard deviation for energy domain parameters for each combination of vowels [a, i, u], consonants [f, v, s] and forms (singleton vs. geminate), averaged over speakers and repetitions (the numerical values are presented in Table XXIX in Appendix).

Figure 8 does not highlight any clear trend for any of the parameters, in particular in relation to the gemination. A statistical analysis based on a multi-factor univariate ANOVA test considering the fixed factors Form, Vowel, Consonant and Gender was thus performed over all combined words. Test results are presented in Table XVI, and show that Form is not a significant factor, since no parameter presented a significant variation of average values caused by gemination. As for the other factors, as expected, Vowel and Gender lead to significant differences for all parameters measured on vowels, while Consonant was the only significant factor for all parameters measured on the consonant.

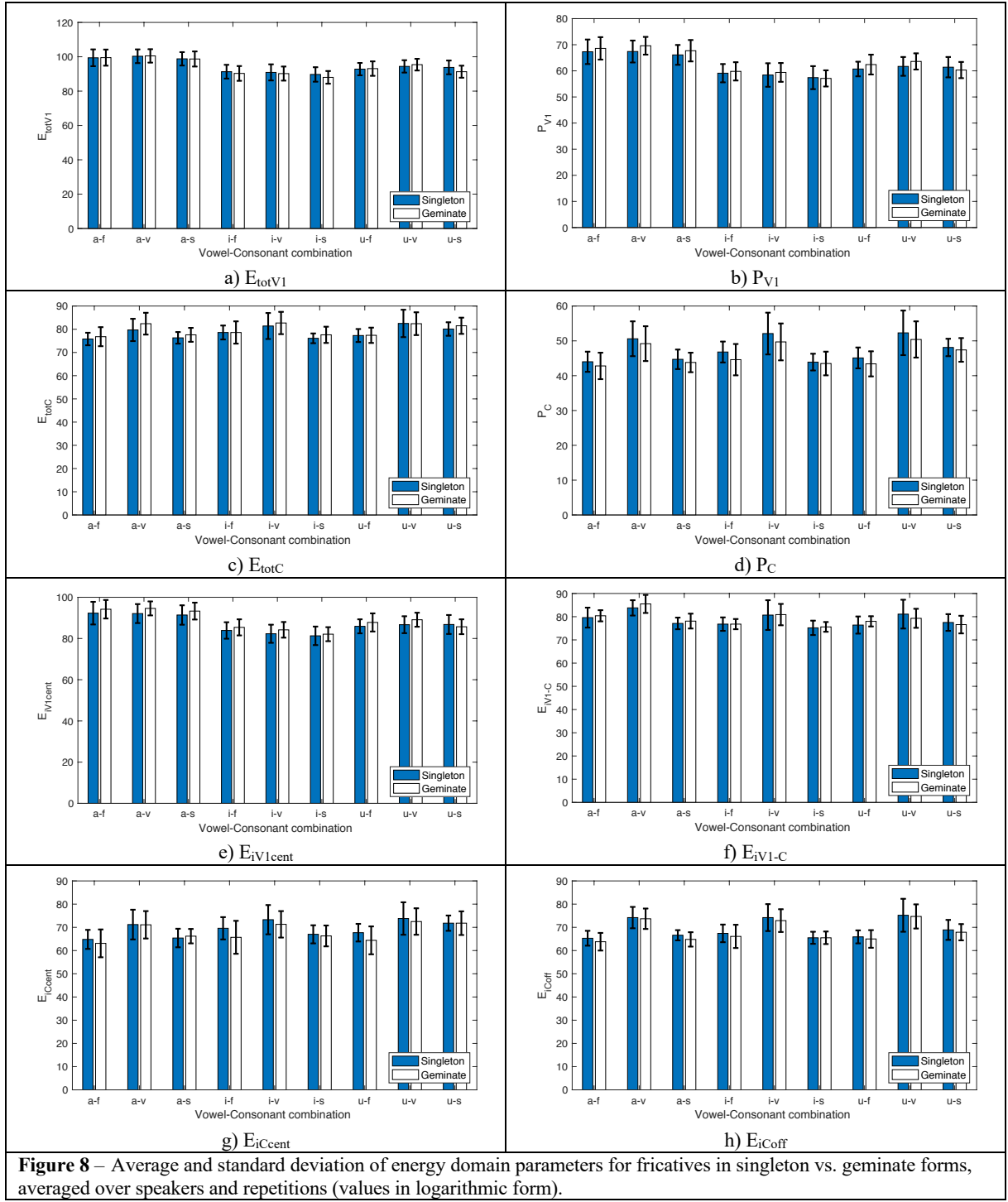

**Figure 8** – Average and standard deviation of energy domain parameters for fricatives in singleton vs. geminate forms, averaged over speakers and repetitions (values in logarithmic form).

|  | $E_{totV1}$ | $P_{mV1}$ | $E_{totC}$ | $P_{mC}$ | $E_{iV1cent}$ | $E_{iV1-C}$ | $E_{iCcent}$ | $E_{iCoffset}$ |
|---|---|---|---|---|---|---|---|---|
| **Form** |  |  |  |  |  |  |  |  |
| **Vowel** | X | X |  |  | X | X | X |  |
| **Consonant** |  |  | X | X |  | X | X | X |
| **Gender** | X | X |  |  | X |  |  |  |

**Table XVI** - Results of the multi-factor univariate ANOVA test performed for fricatives on energy domain parameters using Form, Vowel, Consonant and Gender as fixed factors for all words; a checked cell at the intersection between a parameter and a factor indicates a significant difference between average values for the parameter with respect to the factor.

## 5. Discussion

### 5.1. Effect of gemination on affricates

Results of the analysis presented in Section 4.1.1 showed a significant increase in consonant duration (both closure and fricative sections) and a decrease of pre-consonant vowel duration in geminates vs. singleton affricates. No significant variation was observed in the post-consonant vowel duration. Word duration Utd was only marginally affected by gemination, with significant variations observed only for specific combinations of vowels and consonants, suggesting the existence of a compensation effect between V1d and Cd.

In the frequency domain, F0 significantly increased when moving from singleton to geminate only for male speakers, and only for reference frames related to V1, showing an average increase of about 13 Hz, which is actually a perceptually relevant variation (Hess, 1983). No clear explanation for this variation can be provided, suggesting the need for further research on this specific topic. No significant variations were observed for V1 and V2 formants in any frame for neither female nor male speakers.

Energy parameters were analyzed both separately for voiced and voiceless affricates, and on all affricates combined. The total energy of the consonant $E_{totC}$ was the only parameter that presented significant variations due to gemination in all the three groups, confirming findings already reported for nasal and liquids. Other parameters were significantly affected by gemination in some but not all groups, as detailed in Section 4.1.3.

### 5.2. Effect of gemination on fricatives

Time domain parameters were strongly correlated with gemination. In particular, V1d and Cd were significantly different in singletons vs. geminates, and in particular a longer consonant and a shorter pre-consonant vowel in geminates.

On the other hand, frequency domain parameters were not significantly different in singletons vs. geminates for fricatives: neither pitch F0 nor formants F1, F2 and F3 in both V1 and V2 showed any significant variation with gemination.

Similar results were obtained for energy domain parameters. None of the energy domain parameters was significantly affected by gemination. This finding tells apart fricatives from nasals, liquids and affricates, and recalls the results reported for stops in (Esposito and Di Benedetto, 1999), where no significant variations with gemination were detected for any energy related parameter.

### 5.3. Classification of affricates and fricatives based on durational parameters

As for nasals and liquids in (Di Benedetto and De Nardis, 2020), classification tests of geminate vs. singleton words using time domain parameters as test variables were carried out on affricates and fricatives.

Table XVII shows the classification error percentage for tests using V1d, Cd and V2d for male and female speakers, and for all words combined.

|  |  | V1d | Cd | C1d | C2d | V2d |
|---|---|---|---|---|---|---|
| **Affricates** | Combined | 23.8 | 19.0 | 18.3 | 38.7 | 47.4 |
|  | Male | 21.8 | 17.1 | 14.8 | 42.6 | 47.2 |
|  | Female | 26.9 | 20.4 | 19.4 | 36.1 | 50.5 |
| **Fricatives** | Combined | 17.9 | 12.0 | - | - | 45.1 |
|  | Male | 11.7 | 7.4 | - | - | 40.7 |
|  | Female | 24.1 | 14.8 | - | - | 41.4 |

**Table XVII** – Error classification rate of singleton vs. geminate of affricates and fricatives based on unidimensional MLC tests on time domain parameters V1d, Cd, V2d and, for affricates only, C1d and C2d, for separate female and male speakers, and for all words combined.

Results in Table XVII are in good agreement with the results of the ANOVA tests shown in Section 4. The parameters that varied most significantly due to gemination, that is C1d for affricates and Cd for fricatives, also led to the lowest classification error rates. Classification tests using V1d led to higher error rates, coherently with the weaker significance for V1d variations observed in Section 4.

Additional tests were carried out, to investigate the combination of multiple parameters in the classification of geminate vs. singleton consonants. The analysis focused on the combination of Cd and V1d for both affricates, and fricatives, and of C1d and V1d for affricates. V2d and C2d were not considered based on the high error rates observed in Table XVII when using such parameters. Following the same approach adopted in (Di Benedetto and De Nardis, 2020), parameters were combined in two ways: first, they were used as variables in bidimensional MLC tests; secondly, the ratios Cd/V1d and C1d/V1d (for affricates) were used in unidimensional tests.

Table XVIII shows the classification error rates for the following three cases: 1) female speakers, 2) male speakers and 3) all speakers combined. Results of bidimensional tests indicate that in affricates the introduction of V1d leads to performance improvement in all three cases. In fricatives a performance improvement was observed for male speakers and for all speakers combined, but not for female speakers.

The results of unidimensional tests using the Cd/V1d and C1d/V1d ratios did not consistently lead to a performance improvement in classification, as already observed in (Di Benedetto and De Nardis, 2020) for nasals and liquids. In affricates a slight improvement (less than 1%) was observed for combined male speakers and female speakers when switching from C1d to C1d/V1d, while in fricatives a performance improvement was obtained only for male speakers. In all the other cases, the adoption of the ratio in place of the primary acoustic cue for gemination led to similar or worse classification performance. Note, however, that in affricates all groups show a lower minimum error rate using C1d/V1d rather than Cd/V1d, confirming that in affricates closure duration is a better cue to gemination than consonant duration.

|  |  | Bidimensional | | Unidimensional | |
|---|---|---|---|---|---|
|  |  | (Cd, V1d) | (C1d, V1d) | Cd/V1d | C1d/V1d |
| **Affricates** | Combined | 17.6 | 15.3 | 22.9 | 17.6 |
|  | Male | 15.3 | 10.7 | 20.4 | 15.3 |
|  | Female | 19.0 | 17.6 | 25.0 | 19.0 |
| **Fricatives** | Combined | 10.5 | - | 12.0 | - |
|  | Male | 3.7 | - | 3.1 | - |
|  | Female | 16.7 | - | 21.6 | - |

**Table XVIII** – Error classification rates of singleton vs. geminate for affricates and fricatives in unidimensional MLC tests using ratios Cd/V1d and C1d/V1d (for affricates only) and in bidimensional tests using (Cd, V1d) and (C1d, V1d) (for affricates only), for separate female and male speakers, and for all combined words.

The thresholds on Cd/V1d that led to the best classification performance in the MLC test, corresponding to the Points of Equal Probability (PEPs) between the two Gaussian distributions fitted on singleton vs. geminate data, are presented in Table XIX. Table XIX also presents the thresholds that led to the best classification performance in a heuristic test that explored all possible thresholds, as already analyzed for nasals and liquids in (Di Benedetto and De Nardis, 2020). Table XIX shows that in both tests the best classification performance was obtained for each consonant category with different thresholds for the Cd/V1d ratio; in affricates, in particular, singletons vs. geminates were best classified with a threshold close to 2, while in stops and fricatives thresholds leading to the best classification rate were close to 1.

When considering the C1d/V1d ratio, the thresholds that lead to the best classification rates for affricates are lower; nevertheless, the classification error percentage is still higher than for fricatives, as previously shown in Table XVIII for the MLC test, and as clearly highlighted in Figure 9, presenting the classification error percentage of the heuristic test using the Cd/V1d ratio for fricatives and using both Cd/V1d and C1d/V1d ratio for affricates. Figure 9 also provides further evidence that closure duration leads to better classification performance than consonant duration for gemination in affricates.

|  |  | Cd/V1d threshold | | C1d/V1d threshold | |
|---|---|---|---|---|---|
|  |  | MLC PEP | Heuristic | MLC PEP | Heuristic |
| **Affricates** | Combined | 1.89 | 1.61 | 0.92 | 0.76 |
|  | Male | 1.84 | 1.71 | 0.89 | 0.68 |
|  | Female | 1.92 | 1.44 | 0.95 | 0.80 |
| **Fricatives** | Combined | 1.32 | 1.14 | - | - |
|  | Male | 1.14 | 1.14 | - | - |
|  | Female | 1.45 | 1.14 | - | - |

**Table XIX** – Thresholds for singleton vs. geminate classification in affricates and fricatives using the ratios Cd/V1d and C1d/V1d (for affricates only) for separate female and male speakers, and for all combined words; thresholds were determined both as the Point of Equal Probability (PEP) resulting from the assumption of Gaussian distributions for the two groups of geminate and singleton, and heuristically as the value that minimizes classification errors.

The higher classification error rate in affricates than in fricatives (and even higher than in nasals and liquids, see (Di Benedetto and De Nardis, 2020)) seems to support the hypothesis that some (or all) affricates may not admit both singleton and geminated versions in intervocalic position in Italian, and that therefore the results may reflect the difficulty of the speakers in producing words for which they lack the knowledge of how to express phonetically a phonological element, and therefore produce it in an artificial manner. Muljacic (1972), in particular, suggested that dental affricates may never appear in singleton form when intervocalic.

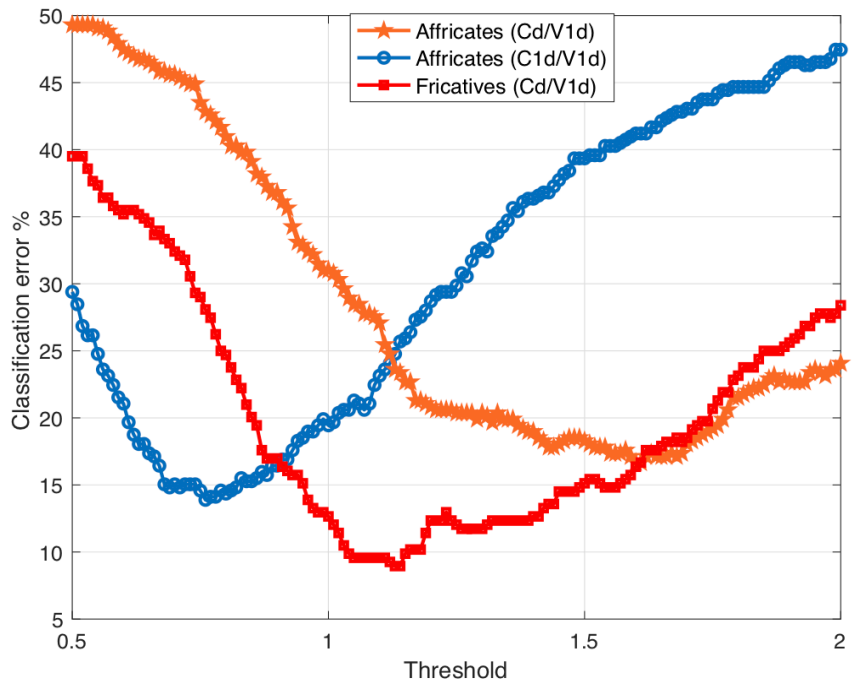

**Figure 9** – Error classification rate in the heuristic test as a function of the Cd/V1d threshold (both affricates and fricatives) and of the C1d/V1d threshold (affricates only).

The four affricates were therefore split into two groups, the non-dental affricates [tʃ, dʒ], vs. the dental affricates [ts, dz]. The heuristic classification test using C1d/V1d as test variable was repeated separately for the two groups; results are presented in Figure 10, that also shows the results for all affricates.

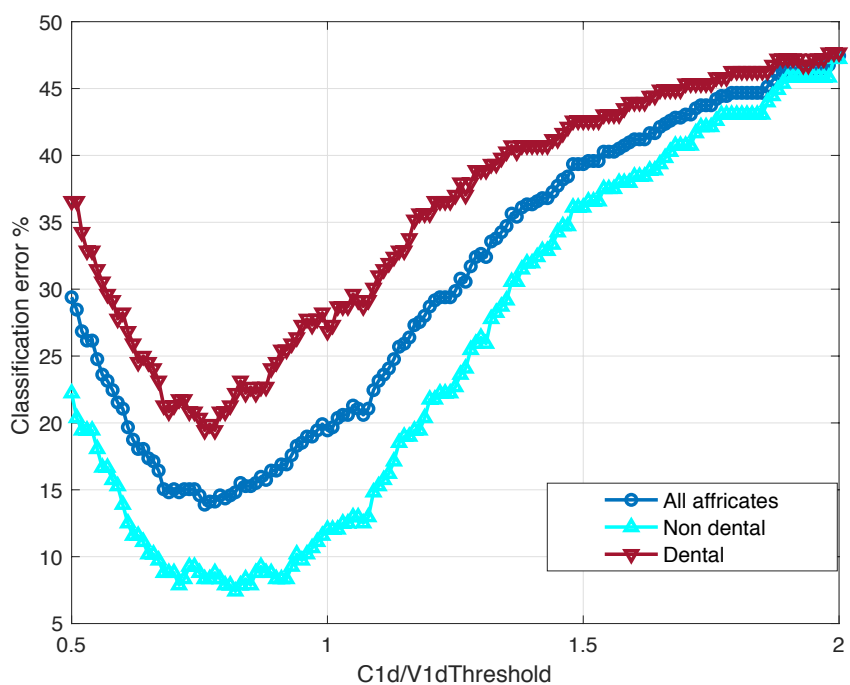

**Figure 10** – Error classification rate in the heuristic test as a function of the C1d/V1d threshold for all affricate consonants considered in this work vs. dental ones ([ts], [dz]) vs. non-dental ones ([tʃ], [dʒ])

Two major observations can be drawn from the results. First, dental affricates are characterized by an error classification rate above 20%, whereas non-dental affricates are affected by an error classification rate lower than 10% for a wide span of C1d/V1d values. This effect is masked when all affricates are combined. This result provides support to Muljacic (1972), that only non-dental affricates actually admit a singleton form in intervocalic position.

Second, the region of C1d/V1d values leading to the lowest classification error percentage is well below 1, and is comparable with the one observed for nasals and liquids in (Di Benedetto and De Nardis, 2020). A full comparison between all consonant categories with respect to gemination will be carried out in the following two subsections.

### 5.4. Comparison of acoustic correlates of gemination for all consonant categories

Results of the present study on affricates and fricatives confirm the observations of previous studies on stops (Esposito and Di Benedetto, 1999) and on nasals and liquids in (Di Benedetto and De Nardis, 2020), and highlight a significant Cd increase in geminate words, compensated by a reduction of the duration of pre-consonant vowel. The lack of a clear impact of gemination of frequency and energy domains parameters also confirms previous studies.

A comparison of the impact of gemination in affricates and fricatives in terms of temporal parameters, vs. nasals and liquids (Di Benedetto and De Nardis, 2020) was carried out. The analysis also included stops, based on the data originally presented in (Esposito and Di Benedetto, 1999) and new statistical analyses on the complete data set of consonants.

Table XX summarizes the average value and standard deviation for the five consonant categories, averaged over all repetitions, speakers, consonants, and vowels. Table XX shows that consonant duration is the parameter showing the largest relative variation across all consonant categories (≈+62% for C1d in affricates, ≈+73% for Cd in fricatives, ≈+133% in nasals, ≈+187% in liquids and ≈+101% for the closure duration Cld in stops) followed by pre-consonant vowel duration V1d (≈-25% in affricates, ≈ -28% in fricatives, ≈-32% in nasals, ≈ -41% in liquids and ≈-26% in stops).

Results of the analysis on the significance of time domain parameter variations for affricates (Table III) and fricatives (Table X) are in good agreement with the analysis carried out in (Esposito and Di Benedetto 1999) for stops and in (Di Benedetto and De Nardis, 2020) for nasals and liquids, although variations of time domain parameters with gemination in affricates are not as sharp as for the other consonant classes. As discussed in Section 5.3, this result can be explained by the bias introduced by dental affricates.

| | | | V1d | Cd | C1d Cld (stops) | C2d | V2d | Utd | Cd/V1d | C1d/V1d Cld/V1d (stops) | C2d/V1d |
|---|---|---|---|---|---|---|---|---|---|---|---|
| **Affricates** | Singleton | Mean | 149.51 | 177.06 | 81.79 | 95.28 | 128.41 | 454.99 | 1.30 | 0.59 | 0.71 |
| | | StD | 33.28 | 43.20 | 25.02 | 40.47 | 27.08 | 41.61 | 0.64 | 0.27 | 0.46 |
| | Geminate | Mean | 111.44 | 254.83 | 133.29 | 121.54 | 125.31 | 491.58 | 2.42 | 1.25 | 1.17 |
| | | StD | 22.48 | 42.67 | 33.03 | 47.44 | 24.12 | 49.02 | 0.77 | 0.43 | 0.59 |
| **Fricatives** | Singleton | Mean | 175.66 | 134.91 | - | - | 118.90 | 429.46 | 0.80 | - | - |
| | | StD | 25.87 | 37.60 | - | - | 25.29 | 45.57 | 0.33 | - | - |
| | Geminate | Mean | 126.58 | 233.25 | - | - | 114.12 | 473.96 | 1.97 | - | - |
| | | StD | 27.14 | 45.07 | - | - | 24.29 | 48.50 | 0.7 | - | - |
| **Nasals** | Singleton | Mean | 183.52 | 90.64 | - | - | 130.05 | 404.20 | 0.51 | - | - |
| | | StD | 27.45 | 14.14 | - | - | 25.43 | 45.07 | 0.12 | - | - |
| | Geminate | Mean | 124.56 | 211.75 | - | - | 124.25 | 460.57 | 1.77 | - | - |
| | | StD | 20.95 | 33.33 | - | - | 25.43 | 43.02 | 0.56 | - | - |
| **Liquids** | Singleton | Mean | 171.92 | 60.56 | - | - | 100.21 | 384.1 | 0.36 | - | - |
| | | StD | 25.75 | 15.33 | - | - | 22.1 | 40.53 | 0.11 | - | - |
| | Geminate | Mean | 121.81 | 174.2 | - | - | 87.74 | 443.86 | 1.52 | - | - |
| | | StD | 27.54 | 28.69 | - | - | 21.45 | 42.87 | 0.51 | - | - |
| **Stops** | Singleton | Mean | 168.33 | 99.8 | 90.79 | - | 145.16 | 413.3 | 0.62 | 0.57 | - |
| | | StD | 28.4 | 22.77 | 19.97 | - | 27.9 | 40.61 | 0.24 | 0.20 | - |
| | Geminate | Mean | 124.4 | 191.46 | 182.05 | - | 137.34 | 453.2 | 1.64 | 1.55 | - |
| | | StD | 25.43 | 46.35 | 36.33 | - | 38.5 | 42.82 | 0.62 | 0.55 | - |

**Table XX** – Average value and standard deviation of the time domain parameters averaged over all the repetitions, speakers, consonants and vowels for affricates, fricatives, nasals, liquids and stops (data for nasals and liquids are from (Di Benedetto and De Nardis, 2020), data for stops are from (Esposito and Di Benedetto 1999) and (GEMMA, 2019)).

Comparison in terms of Spearman Rank correlation required to carry out the test on stops, since this was not originally reported in (Esposito and Di Benedetto, 1999). Table XXI presents the results of the test. A comparison with affricates (Table V) and fricatives (Table XII) shows that both fricatives and stops present a high negative correlation between V1d and Cd (< -0.7); a negative correlation, apparently weaker, was observed for affricates between V1d and C1d (-0.47), and V1d and C2d (-0.47), although when considering the correlation between V1d and Cd=C1d+C2d a stronger effect was observed (-0.7). These results are well in line with those obtained for nasals and liquids, presented in (Di Benedetto and De Nardis, 2020).

| | Singleton | | | Geminate | | |
|---|---|---|---|---|---|---|
| | V1d s. | Cd s. | V2d s. | V1d g. | Cd g. | V2d g. |
| V1d s. | 1.00 | -0.42 | 0.3 | | | |
| Cd s. | -0.42 | 1.00 | -0.25 | | not significant | |
| V2d s. | 0.3 | -0.25 | 1.00 | | | |
| V1d g. | | | | 1.00 | -0.39 | 0.37 |
| Cd g. | | not significant | | -0.39 | 1.00 | -0.34 |
| V2d g. | | | | 0.37 | -0.34 | 1.00 |

| | V1d | Cd | V2d |
|---|---|---|---|
| V1d | 1.00 | -0.72 | 0.37 |
| Cd | -0.72 | 1.00 | -0.31 |
| V2d | 0.37 | -0.31 | 1.00 |

a) Separate groups (singleton vs. geminate)      b) Combined

**Table XXI** - Spearman Rank Correlation Coefficient $r_s$ of time domain parameters for singleton and geminate stop words separately (Table XXIa)) and on all words, singleton and geminate combined (Table XXIb)). Bold characters indicate significant correlations, with threshold set as p*=0.05.

### 5.5. Classification of geminate vs. singleton words across consonant classes

The results on classification of nasals and liquids in (Di Benedetto and De Nardis, 2020), combined with those presented in Section 5.3 for affricates and fricatives formed the basis for a comparison in terms of classification geminate vs. singleton words using time domain parameters as test variables between different consonant classes. Table XXII introduces the results of tests on stops, which were re-analyzed since the classification tests presented in (Esposito and Di Benedetto, 1999) were in a preliminary form; in particular they only focused on closure duration rather than Cd, while tests using V1d were only performed for all words combined.

|  |  | Unidimensional |  |  |  |  |  | Bidimensional |  |
|---|---|---|---|---|---|---|---|---|---|
|  |  | V1d | Cd | Cld | V2d | Cd/V1d | Cld/V1d | (Cd, V1d) | (Cld, V1d) |
| Stops | Combined | 20.2 | 5.6 | 4.0 | 44.1 | 8.3 | 8.2 | 6.0 | 4.2 |
|  | Male | 13.6 | 2.8 | 3.1 | 46.6 | 3.1 | 2.2 | 2.5 | 2.8 |
|  | Female | 25.9 | 7.7 | 5.9 | 43.2 | 14.2 | 13.3 | 7.7 | 4.6 |

**Table XXII** – Classification error rate of singleton *vs.* geminate stop consonants based on unidimensional MLC tests on time domain parameters V1d, Cd, Cld, V2d, Cd/V1d and Cld/V1d, and on bidimensional MLC tests on (Cd, V1d) and (Cld, V1d), for separate female and male speakers groups, and for combined groups.

Results in Table XXII show that in stops, as it was the case for affricates, closure duration is the most relevant parameter for characterizing gemination; consonant duration Cd led in fact to slightly worse performance, with a 5.6% of errors vs. 4.0%. The same remark holds for the other test variables based on Cld vs. Cd, both in unidimensional and bidimensional tests. The results for stops show that the introduction of V1d does not lead to any performance improvement when all combined words are considered, confirming the results presented in (Esposito and Di Benedetto, 1999) for bidimensional tests using Cld and V1d.

The thresholds on Cd/V1d and Cld/V1d that led to the best classification performance for stops in the MLC test and in the heuristic test introduced in (Di Benedetto and De Nardis, 2020), and previously performed on the other consonant classes, are presented in Table XXIII.

|  |  | Cd/V1d threshold |  | Cld/V1d threshold |  |
|---|---|---|---|---|---|
|  |  | MLC PEP | Heuristic | MLC PEP | Heuristic |
| Stops | Combined | 1.02 | 1.02 | 0.93 | 0.85 |
|  | Male | 0.93 | 1.02 | 0.86 | 0.85 |
|  | Female | 1.09 | 0.77 | 0.99 | 0.71 |

**Table XXIII** – Thresholds for singleton *vs.* geminate classification in stop consonants using the ratios Cd/V1d and Cld/V1d for separate female and male speakers, and for all combined words; thresholds were determined both as the Point of Equal Probability (PEP) resulting from the assumption of Gaussian distributions for the two groups of geminate and singleton words, and heuristically as the value that minimizes classification errors.

These results can be combined with those reported in (Di Benedetto and De Nardis, 2020) for nasals and liquids and in Section 5.3 for affricates and fricatives so to understand whether the Cd/V1d ratio is an invariant property across consonant classes, in analogy to its invariance across speaking rates, suggested in (Pickett *et al.*, 1999). If this were the case, the ratio between the primary acoustic cue, that is closure duration in stops and affricates and Cd in all other consonant classes, and V1d, could be used as a test variable in the classification of singletons vs. geminates for all consonant classes. In light of the analysis on dental vs. non-dental affricates carried out in Section 5.3, only non-dental affricates will be considered in the following.

Results of Table XVIII and Table XXII confirm the observation on nasals and liquids (Di Benedetto and De Nardis, 2020), that the ratio does not lead in general to better classification rates than the primary cue alone. Furthermore, results presented in Table XIX and Table XXIII, combined with the results on nasals and liquids presented in Table XIX in (Di Benedetto and De Nardis, 2020) show that best classification performance is achieved for the five consonant categories for different ratio threshold values, ranging from 0.74 in liquids to 1.14 in fricatives. It was also shown, however, that similar error rates were achieved for a wide range of ratio threshold values, making it possible to identify a favourable threshold across consonant classes, at the expense of a small performance loss with respect to the best threshold of each class.

In order to simplify the notation, we will indicate from now on the consonant clue, that is Cd for fricatives, nasals, and liquids, vs. Cld for stops and C1d for affricates, with C.

Classification tests were therefore performed on the combined set of all consonants. Table XXIV shows the classification error rate obtained by unidimensional MLC tests using V1d, C, V2d, Utd and the ratio C/V1d, as well as in a bidimensional tests using (C, V1d). Tests were performed on combined words and on male speakers and female speakers separately. In unidimensional tests C and C/V1d were the parameters leading to the best performance: C/V1d minimized the error percentage for combined words and male speakers, while C led to the best results for female speakers. It should be noted that the bidimensional test led to best classification rates for all groups, suggesting that the use of a secondary gemination cue V1d may lead to improved classification rates.

|  |  | Unidimensional |  |  |  |  | Bidimensional |
|---|---|---|---|---|---|---|---|
|  |  | V1d | C | V2d | Utd | C/V1d | (C, V1d) |
| All consonants | Combined | 18.9 | 10.9 | 44.3 | 32.8 | 7.8 | 7.4 |
|  | Male | 14.1 | 10.3 | 44.1 | 34.8 | 3.7 | 3.6 |
|  | Female | 23.8 | 11.6 | 43.8 | 30.4 | 13.6 | 10.9 |

**Table XXIV** – Classification error rate of singleton vs. geminate consonants, for all consonants combined, obtained with unidimensional MLC tests using V1d, C, V2d, Utd, and C/V1d, and in a bidimensional test using (C, V1d), for female and male speakers separately, and for all combined words.

The heuristic test, as described in Section 5.5, was applied as well to all consonants, again for all words combined and for male and female speakers separately. Results of the tests are presented in Table XXV, showing that error rates are in this case as well lower than with MLC tests.

|  |  | Unidimensional |  |  |  |  |
|---|---|---|---|---|---|---|
|  |  | V1d | C | V2d | Utd | C/V1d |
| All consonants | Combined | 18.4 | 10.1 | 43.0 | 32.4 | 7.5 |
|  | Male | 12.6 | 9.4 | 42.8 | 32 | 3.0 |
|  | Female | 22.8 | 10.5 | 42.8 | 30.1 | 11.6 |

**Table XXV** – Error classification rates of singleton vs. geminate for all consonants combined in unidimensional heuristic tests using the V1d, C, V2d, Utd, and C/V1d for separate female and male speakers, and for all combined words.

In particular, tests using the C/V1d ratio led to improved classification rates for combined and male speakers, suggesting that C/V1d may be a valid classification parameter. The adoption of a common C/V1d threshold across all consonant classes slightly increases, as expected, the classification error percentage for each of the consonant classes. This phenomenon is highlighted in Figure 11, showing the error classification percentage as a function of the threshold for each class, for all consonants combined.

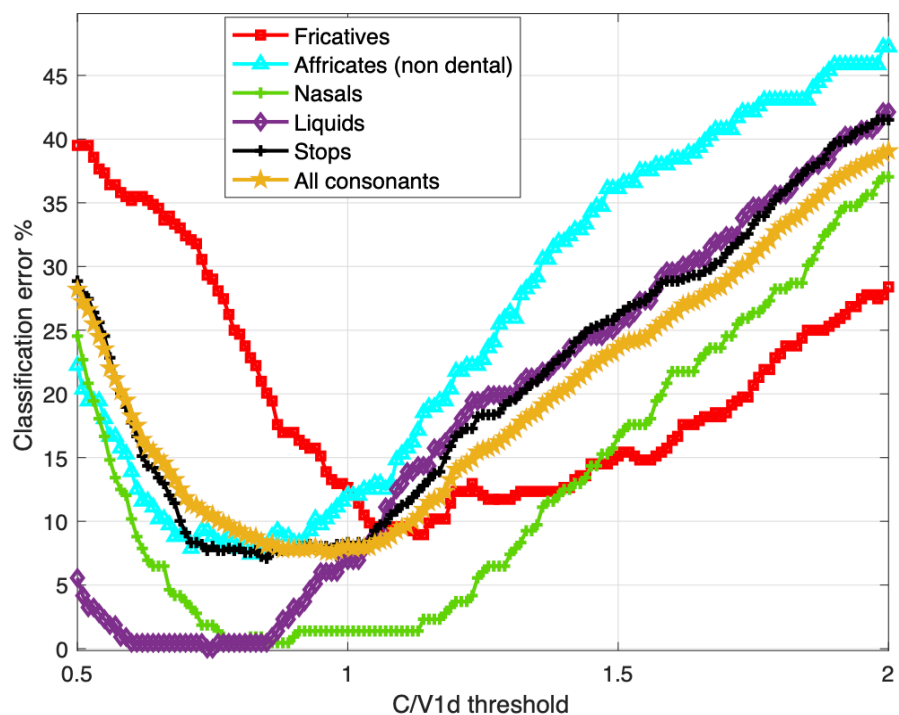

**Figure 11** – Error classification rate using the heuristic test as a function of the C/V1d threshold for each consonant class and for all words combined, with and without affricates (data for nasals and liquids are taken from Figure 8 in (Di Benedetto and De Nardis, 2020).)

Figure 11 also highlights that the best threshold value is somewhat lower but close to 1 for liquids, nasals and stops, and somewhat larger but close to 1 for fricatives.

Overall, classification based on C/V1d leads to an excellent classification performance, with error percentages that are below 10%, as also shown in Table XXV.

## 6. Conclusions

This research investigated the impact of gemination on affricate and fricative Italian consonants, based on acoustic analyses of disyllabic words (VCV vs. VCCV) in a symmetrical context of cardinal vowels [a, i, u]. These words belong to the GEMMA project database (GEMMA, 2019). Time domain, frequency domain and energy domain measurements were collected in different frames within the word, corresponding to crucial events such as vowel-to-consonant transition and vowel and consonant stable portions.

The most relevant outcomes can be summarized as follows:
- a general tendency of shortening the pre-consonant vowel and of lengthening the consonant in a geminate word, that was observed in previous studies for stops (Esposito and Di Benedetto, 1999) and for nasals and liquids (Di Benedetto and De Nardis, 2020), was confirmed for both affricates and fricatives;
- differently from what was observed for nasals and liquids in (Di Benedetto and De Nardis, 2020), a degree of correlation between the two aforementioned effects was also observed in singletons *vs.* singletons words. It is important however to point out that such correlation is stronger in geminates vs. geminates, and even more in geminates *vs.* singletons, again confirming the hypothesis by Shrotiya *et al.*, (1995) on the preservation of rhythmical structures;
- the analysis of pitch and formants did not highlight any systematic effect of gemination, with exception of an increased F0 for V1 in male speakers for affricates, as it was observed for nasals in (Di Benedetto and De Nardis, 2020). No clear explanation for this phenomenon was found, requiring thus further investigations.
- no significant energy variations were observed in previous studies for stops, while a mild variation in consonant energy was observed for liquids and nasals. The present study confirms that energy parameters are only weakly affected by gemination. No variations were in fact observed in fricatives, while a slight tendency to emphasize both energy and power of the geminate utterance emerged for affricates, even if differences are limited to a few dBs;
- the use of the primary acoustic cue for classification of singletons vs. geminates led to the best classification rates for both affricates and fricatives. A slight performance improvement was achieved in both affricates and fricatives classification rate by combining the primary cue with first vowel duration V1d in a bidimensional classifier;
- the C/V1d ratio (ratio between the consonantal durational clue i.e. consonant duration Cd for fricatives, nasals, and liquids, vs. consonant closure duration C1d for stops and affricates) was investigated as an across-consonant parameter for detecting gemination; results highlighted that although the optimal C/V1d threshold varies across different consonant classes, a classification for all combined consonants except affricates achieves its optimal performance for a threshold value of about 1;
- a detailed analysis on affricates, carried out dividing them in dental vs. non-dental, highlighted that gemination in non-dental affricates is easier to detect, with classification performance comparable to the one obtained for other consonant classes, while dental affricates lead to poor classification performance; this finding supports the hypothesis of Muljacic (1972) that dental affricates do not admit singleton forms in intervocalic position.

# 7. Appendix – Average value and standard deviation of time domain and energy domain parameters

## 7.1. Affricates

|   |        | V1d (msecs) | | C1d (msecs) | | C2d (msecs) | | Cd (msecs) | | V2d (msecs) | | Utd (msecs) | |
|---|--------|------|------|------|------|------|------|------|------|------|------|------|------|
|   |        | Mean | StD  | Mean | StD  | Mean | StD  | Mean | StD  | Mean | StD  | Mean | StD  |
| a | atʃa   | 160.0 | 27.6 | 73.1  | 34.7 | 100.9 | 20.5 | 174.0 | 31.8 | 112.3 | 19.6 | 446.3 | 43.8 |
|   | attʃa  | 113.2 | 19.2 | 137.8 | 13.9 | 128.7 | 28.1 | 266.5 | 30.4 | 107.5 | 12.2 | 487.2 | 29.3 |
|   | adʒa   | 169.0 | 20.6 | 92.0  | 18.9 | 49.1  | 13.6 | 141.0 | 27.0 | 142.3 | 26.1 | 452.3 | 47.4 |
|   | addʒa  | 127.3 | 16.0 | 156.1 | 17.7 | 61.5  | 11.0 | 217.6 | 24.1 | 125.9 | 15.9 | 470.9 | 42.2 |
|   | atsa   | 121.3 | 23.3 | 89.6  | 11.0 | 129.8 | 34.0 | 219.4 | 36.0 | 109.9 | 23.1 | 450.6 | 37.0 |
|   | attsa  | 106.0 | 18.7 | 112.2 | 18.8 | 167.0 | 22.0 | 279.2 | 32.7 | 117.4 | 20.6 | 502.6 | 43.5 |
|   | adza   | 163.4 | 24.7 | 89.9  | 13.5 | 78.6  | 19.3 | 168.5 | 22.5 | 139.7 | 18.9 | 471.7 | 42.9 |
|   | addza  | 127.8 | 24.5 | 139.8 | 35.3 | 102.3 | 23.0 | 242.1 | 34.0 | 136.3 | 29.0 | 506.2 | 57.4 |
| i | itʃi   | 137.4 | 20.8 | 64.0  | 29.2 | 122.4 | 16.2 | 186.4 | 35.3 | 104.6 | 17.9 | 428.4 | 29.8 |
|   | ittʃi  | 99.3  | 17.9 | 122.8 | 20.4 | 158.4 | 26.1 | 281.2 | 31.9 | 110.7 | 21.0 | 491.3 | 37.5 |
|   | idʒi   | 166.7 | 28.3 | 95.9  | 17.5 | 52.6  | 15.7 | 148.5 | 25.1 | 141.6 | 30.6 | 456.8 | 53.4 |
|   | iddʒi  | 111.7 | 21.3 | 162.1 | 28.2 | 74.1  | 25.5 | 236.2 | 41.2 | 129.4 | 30.6 | 477.3 | 56.6 |
|   | itsi   | 106.7 | 25.9 | 84.4  | 20.2 | 149.6 | 31.3 | 234.0 | 39.6 | 109.7 | 18.1 | 450.4 | 32.2 |
|   | ittsi  | 94.5  | 17.9 | 114.0 | 31.4 | 171.0 | 34.7 | 285.0 | 37.6 | 123.2 | 22.8 | 502.7 | 48.0 |
|   | idzi   | 148.4 | 37.5 | 85.9  | 16.5 | 90.9  | 21.6 | 176.8 | 30.9 | 148.1 | 20.7 | 473.4 | 35.7 |
|   | iddzi  | 104.7 | 23.9 | 136.5 | 36.4 | 120.2 | 38.1 | 256.7 | 42.3 | 139.7 | 19.0 | 501.1 | 53.0 |
| u | utʃu   | 163.6 | 27.4 | 66.0  | 37.9 | 103.7 | 24.0 | 169.8 | 34.4 | 131.7 | 23.7 | 465.0 | 32.0 |
|   | uttʃu  | 110.9 | 25.4 | 151.1 | 39.4 | 123.0 | 24.7 | 274.1 | 48.1 | 125.0 | 22.4 | 509.9 | 51.7 |
|   | udʒu   | 173.5 | 32.1 | 85.7  | 21.1 | 44.1  | 16.5 | 129.9 | 27.2 | 146.1 | 26.5 | 449.5 | 45.0 |
|   | uddʒu  | 120.2 | 21.6 | 154.0 | 21.3 | 61.3  | 20.8 | 215.3 | 32.0 | 137.3 | 29.9 | 472.8 | 67.7 |
|   | utsu   | 133.2 | 30.6 | 73.3  | 26.9 | 140.7 | 22.4 | 214.0 | 32.6 | 115.3 | 16.3 | 462.5 | 41.1 |
|   | uttsu  | 103.8 | 21.9 | 96.3  | 20.4 | 178.8 | 19.4 | 275.0 | 23.6 | 115.1 | 15.8 | 493.9 | 40.4 |
|   | udzu   | 150.8 | 23.7 | 81.6  | 18.8 | 80.9  | 18.1 | 162.5 | 29.2 | 139.7 | 23.8 | 453.0 | 44.8 |
|   | uddzu  | 117.7 | 17.1 | 116.8 | 26.9 | 112.3 | 29.4 | 229.0 | 42.7 | 136.4 | 20.0 | 483.1 | 43.1 |

**Table XXVI** - Average and standard deviation of time domain parameters for affricate words in singleton vs. geminate forms, averaged over all repetitions and speakers (all values are expressed in milliseconds).

|   |        |      | $E_{totV1}$ | $P_{V1}$ | $E_{totC1}$ | $P_{C1}$ | $E_{totC2}$ | $P_{C2}$ | $E_{totC}$ | $P_C$ | $E_{iV1cent}$ | $E_{iV1-C1}$ | $E_{iC1cent}$ | $E_{iC1-C2}$ | $E_{iC2cent}$ | $E_{iC2off}$ |
|---|--------|------|------|------|------|------|------|------|------|------|------|------|------|------|------|------|
| a | atʃa   | Mean | 99.7 | 67.8 | 75.5 | 47.6 | 83.1 | 53.2 | 83.9 | 61.7 | 92.5 | 81.1 | 64.8 | 71.1 | 78.9 | 72.1 |
|   |        | Std  | 3.2  | 2.9  | 2.7  | 4.1  | 3.1  | 3.1  | 2.5  | 2.7  | 3.1  | 2.7  | 9.0  | 6.5  | 3.1  | 2.9  |
|   | atʃtʃa | Mean | 99.2 | 68.8 | 74.9 | 43.4 | 85.9 | 55.3 | 86.4 | 62.2 | 94.0 | 82.7 | 54.1 | 67.3 | 81.6 | 73.1 |
|   |        | Std  | 4.0  | 3.5  | 3.5  | 3.5  | 2.8  | 3.0  | 2.7  | 2.7  | 4.0  | 3.3  | 4.1  | 5.9  | 2.9  | 3.2  |
|   | adʒa   | Mean | 99.6 | 67.2 | 80.7 | 51.2 | 79.5 | 52.7 | 83.8 | 62.3 | 92.1 | 84.1 | 74.8 | 73.3 | 75.9 | 77.3 |
|   |        | Std  | 3.2  | 3.4  | 4.2  | 4.5  | 3.0  | 3.4  | 2.8  | 2.8  | 3.6  | 3.4  | 4.9  | 4.4  | 3.7  | 4.3  |
|   | adʒdʒa | Mean | 100.5| 69.5 | 82.8 | 51.0 | 81.4 | 53.4 | 85.9 | 62.5 | 94.6 | 85.6 | 72.1 | 72.3 | 77.0 | 78.6 |
|   |        | Std  | 3.7  | 3.5  | 3.9  | 4.0  | 3.5  | 3.9  | 2.5  | 2.5  | 3.6  | 3.0  | 3.8  | 3.9  | 4.1  | 4.3  |
|   | atsa   | Mean | 97.2 | 66.2 | 73.2 | 43.7 | 71.7 | 40.7 | 75.9 | 52.7 | 91.2 | 80.1 | 58.2 | 61.7 | 63.9 | 66.4 |
|   |        | Std  | 4.6  | 4.0  | 4.7  | 4.7  | 1.7  | 2.5  | 2.7  | 3.1  | 4.2  | 3.5  | 6.3  | 4.9  | 3.4  | 2.2  |
|   | atstsa | Mean | 99.0 | 68.8 | 72.9 | 42.6 | 73.8 | 41.6 | 76.5 | 52.1 | 93.8 | 82.9 | 54.9 | 60.8 | 64.2 | 69.4 |
|   |        | Std  | 4.0  | 3.5  | 3.2  | 3.2  | 2.5  | 2.5  | 2.7  | 2.6  | 3.7  | 4.5  | 4.0  | 4.3  | 3.0  | 3.5  |
|   | adza   | Mean | 99.8 | 67.8 | 79.7 | 50.3 | 77.4 | 48.5 | 82.1 | 59.9 | 92.5 | 83.4 | 71.8 | 67.4 | 70.3 | 75.5 |
|   |        | Std  | 4.8  | 4.6  | 4.6  | 4.6  | 3.9  | 4.3  | 3.7  | 4.0  | 5.1  | 3.7  | 6.3  | 6.3  | 5.1  | 3.8  |
|   | adzdza | Mean | 100.2| 69.1 | 81.2 | 49.7 | 80.4 | 50.3 | 84.1 | 60.3 | 93.9 | 85.4 | 71.3 | 70.2 | 70.5 | 78.0 |
|   |        | Std  | 3.6  | 3.4  | 4.9  | 4.5  | 4.0  | 4.2  | 4.2  | 3.8  | 3.7  | 3.1  | 4.9  | 5.3  | 6.0  | 4.3  |
| i | itʃi   | Mean | 89.9 | 58.7 | 70.4 | 43.1 | 85.2 | 54.4 | 85.4 | 62.9 | 83.4 | 76.1 | 63.5 | 69.2 | 80.8 | 71.6 |
|   |        | Std  | 5.2  | 4.9  | 5.5  | 6.6  | 2.5  | 2.7  | 2.4  | 2.8  | 4.7  | 4.1  | 9.2  | 5.8  | 2.9  | 3.6  |
|   | itʃtʃi | Mean | 89.0 | 59.3 | 72.5 | 41.7 | 87.3 | 55.6 | 87.4 | 63.0 | 84.3 | 76.6 | 56.0 | 66.2 | 81.4 | 73.9 |
|   |        | Std  | 3.6  | 3.4  | 4.2  | 4.6  | 3.0  | 3.4  | 2.9  | 3.2  | 3.3  | 3.3  | 5.6  | 6.2  | 4.5  | 4.9  |

|   |   |      | | | | | | | | | | | | | |
|---|---|------|---|---|---|---|---|---|---|---|---|---|---|---|---|
|   | idʒi | Mean | 91.1 | 58.9 | 78.7 | 48.8 | 79.2 | 52.3 | 82.9 | 61.3 | 83.9 | 77.7 | 71.3 | 72.8 | 76.8 | 76.9 |
|   |      | Std  | 3.9 | 3.7 | 6.1 | 6.3 | 2.8 | 2.4 | 2.8 | 3.1 | 4.0 | 5.0 | 6.2 | 3.5 | 2.2 | 2.5 |
|   | idʒdʒi | Mean | 88.8 | 58.4 | 80.1 | 48.0 | 80.6 | 52.3 | 84.1 | 60.4 | 83.4 | 77.7 | 70.0 | 72.2 | 76.4 | 77.3 |
|   |      | Std  | 3.6 | 3.1 | 5.1 | 5.2 | 3.2 | 2.8 | 2.9 | 3.2 | 2.9 | 3.7 | 6.7 | 3.6 | 3.0 | 3.7 |
|   | itsi | Mean | 87.8 | 57.6 | 73.2 | 43.9 | 74.2 | 42.7 | 77.3 | 53.7 | 82.3 | 77.4 | 59.3 | 61.3 | 66.4 | 66.3 |
|   |      | Std  | 4.8 | 4.1 | 4.5 | 4.9 | 3.0 | 2.6 | 2.9 | 3.0 | 4.1 | 4.9 | 8.3 | 4.6 | 2.5 | 3.0 |
|   | itstsi | Mean | 89.3 | 59.5 | 72.3 | 41.8 | 74.6 | 42.4 | 77.4 | 52.9 | 84.7 | 78.2 | 56.8 | 62.0 | 66.6 | 66.2 |
|   |      | Std  | 4.1 | 3.8 | 6.3 | 6.7 | 2.5 | 2.2 | 3.3 | 3.4 | 3.8 | 4.5 | 7.2 | 4.9 | 3.1 | 2.7 |
|   | idzi | Mean | 88.0 | 56.7 | 76.9 | 47.8 | 76.2 | 46.8 | 79.9 | 57.5 | 80.9 | 76.9 | 70.7 | 68.5 | 69.0 | 72.8 |
|   |      | Std  | 4.2 | 3.8 | 6.0 | 5.9 | 5.8 | 5.6 | 5.8 | 5.6 | 4.0 | 5.5 | 6.7 | 6.3 | 6.4 | 5.4 |
|   | idzdzi | Mean | 90.6 | 60.6 | 79.6 | 48.4 | 78.6 | 48.0 | 82.6 | 58.6 | 85.4 | 79.9 | 70.9 | 69.1 | 70.2 | 74.9 |
|   |      | Std  | 3.4 | 2.7 | 3.9 | 3.8 | 5.6 | 5.3 | 4.4 | 4.5 | 2.8 | 4.4 | 5.1 | 5.4 | 6.3 | 5.0 |
| u | utʃu | Mean | 93.8 | 61.8 | 72.3 | 45.0 | 81.5 | 51.6 | 82.7 | 60.5 | 87.1 | 77.1 | 65.8 | 71.2 | 77.2 | 70.4 |
|   |      | Std  | 3.7 | 3.3 | 6.1 | 7.2 | 2.8 | 2.9 | 2.1 | 2.1 | 3.9 | 4.6 | 10.6 | 5.7 | 2.9 | 4.0 |
|   | utʃtʃu | Mean | 93.5 | 63.2 | 72.1 | 40.4 | 84.0 | 53.3 | 84.4 | 60.1 | 88.6 | 78.2 | 53.4 | 68.3 | 78.9 | 74.4 |
|   |      | Std  | 2.8 | 2.2 | 3.7 | 3.9 | 2.3 | 2.4 | 2.1 | 2.3 | 2.8 | 3.1 | 3.6 | 5.6 | 2.5 | 4.2 |
|   | udʒu | Mean | 94.9 | 62.6 | 79.1 | 50.0 | 78.1 | 52.1 | 82.2 | 61.1 | 87.1 | 80.5 | 73.1 | 73.7 | 75.6 | 76.7 |
|   |      | Std  | 3.6 | 3.2 | 4.7 | 5.1 | 2.9 | 3.2 | 3.0 | 3.3 | 3.7 | 4.6 | 6.2 | 3.8 | 3.2 | 3.0 |
|   | udʒdʒu | Mean | 93.9 | 63.2 | 80.7 | 48.8 | 80.2 | 52.6 | 84.1 | 60.8 | 88.7 | 80.9 | 70.4 | 72.9 | 75.9 | 77.8 |
|   |      | Std  | 3.1 | 2.4 | 6.1 | 6.1 | 2.8 | 2.9 | 4.0 | 4.3 | 2.3 | 4.0 | 7.5 | 3.7 | 3.0 | 3.7 |
|   | utsu | Mean | 90.2 | 59.1 | 71.9 | 43.7 | 77.3 | 45.9 | 79.0 | 55.7 | 83.2 | 77.4 | 62.2 | 64.6 | 70.7 | 68.9 |
|   |      | Std  | 5.2 | 4.6 | 4.6 | 5.6 | 2.5 | 2.6 | 2.1 | 1.7 | 5.4 | 6.3 | 7.2 | 4.5 | 3.1 | 3.5 |
|   | utstsu | Mean | 91.8 | 61.7 | 74.1 | 44.3 | 81.7 | 49.1 | 83.1 | 58.7 | 86.6 | 80.4 | 58.3 | 65.5 | 74.0 | 71.5 |
|   |      | Std  | 4.5 | 4.0 | 4.6 | 5.0 | 3.8 | 3.7 | 3.1 | 3.1 | 4.4 | 4.5 | 6.7 | 5.2 | 4.1 | 4.7 |
|   | udzu | Mean | 92.6 | 60.8 | 78.9 | 49.7 | 77.5 | 48.7 | 81.6 | 59.6 | 84.9 | 81.4 | 72.0 | 68.3 | 72.1 | 74.5 |
|   |      | Std  | 4.4 | 4.2 | 3.6 | 3.6 | 3.8 | 3.4 | 3.1 | 3.0 | 5.0 | 4.0 | 4.2 | 5.9 | 3.4 | 3.4 |
|   | udzdzu | Mean | 93.1 | 62.5 | 79.9 | 49.4 | 82.1 | 51.8 | 84.4 | 61.0 | 87.2 | 82.3 | 70.3 | 71.7 | 74.7 | 78.7 |
|   |      | Std  | 3.8 | 3.4 | 5.1 | 5.3 | 3.6 | 3.8 | 3.9 | 4.2 | 3.7 | 4.4 | 6.9 | 4.9 | 4.4 | 3.8 |

**Table XXVII** - Average and standard deviation of energy domain parameters for each combination of consonants [tʃ, dʒ, ts, dz], vowels [a, i, u] and singleton vs. geminate form, averaged over repetitions and speakers (values are in logarithmic form; for a list of parameters refer to Section 3.1.4).

## 7.2. Fricatives

|   |      | V1d (msecs) | | Cd (msecs) | | V2d (msecs) | | Utd (msecs) | |
|---|------|------|------|------|------|------|------|------|------|
|   |      | Mean | StD  | Mean | StD  | Mean | StD  | Mean | StD  |
| a | afa  | 165.8 | 18.8 | 151.7 | 21.4 | 111.6 | 28.6 | 429.0 | 37.0 |
|   | affa | 123.2 | 18.1 | 248.3 | 30.3 | 109.1 | 22.4 | 480.7 | 45.6 |
|   | ava  | 188.7 | 21.4 | 83.3  | 13.5 | 123.0 | 26.0 | 395.0 | 45.0 |
|   | avva | 126.5 | 20.7 | 205.8 | 27.0 | 108.0 | 16.2 | 440.3 | 36.6 |
|   | asa  | 175.9 | 17.0 | 147.2 | 13.9 | 122.6 | 27.3 | 445.8 | 37.1 |
|   | assa | 125.3 | 20.4 | 250.1 | 35.6 | 113.7 | 24.3 | 489.2 | 41.0 |
| i | ifi  | 164.3 | 23.1 | 153.0 | 30.7 | 109.9 | 22.9 | 427.3 | 36.4 |
|   | iffi | 115.1 | 27.8 | 253.5 | 37.4 | 112.1 | 27.0 | 480.7 | 49.2 |
|   | ivi  | 185.0 | 26.6 | 90.3  | 13.1 | 118.8 | 23.6 | 394.1 | 48.7 |
|   | ivvi | 122.4 | 28.8 | 202.1 | 29.6 | 117.5 | 29.3 | 442.0 | 61.5 |
|   | isi  | 175.5 | 21.6 | 164.2 | 29.2 | 115.1 | 23.8 | 454.8 | 34.6 |
|   | issi | 124.7 | 27.3 | 260.0 | 36.1 | 113.6 | 20.0 | 498.3 | 44.9 |
| u | ufu  | 163.8 | 34.7 | 163.7 | 26.2 | 118.4 | 21.6 | 446.0 | 43.1 |
|   | uffu | 120.3 | 28.8 | 253.2 | 38.1 | 109.8 | 18.4 | 483.2 | 37.1 |
|   | uvu  | 188.2 | 34.0 | 105.4 | 19.1 | 135.2 | 23.4 | 428.8 | 49.1 |
|   | uvvu | 156.4 | 29.0 | 171.3 | 34.3 | 134.7 | 27.2 | 462.4 | 48.1 |
|   | usu  | 173.8 | 18.7 | 155.3 | 25.8 | 115.4 | 25.5 | 444.5 | 40.7 |
|   | ussu | 125.2 | 25.2 | 255.0 | 39.6 | 108.6 | 24.2 | 488.8 | 39.5 |

**Table XXVIII** - Average value and standard deviation (in milliseconds) of V1d, Cd, V2d and Utd for words containing fricatives, averaged over all repetitions and speakers.

|   |   |   | $E_{totV1}$ | $P_{V1}$ | $E_{totC}$ | $P_C$ | $E_{iV1cent}$ | $E_{iV1-C}$ | $E_{iCcent}$ | $E_{iCoff}$ |
|---|---|---|---|---|---|---|---|---|---|---|
| a | afa | Mean | 99.4 | 67.3 | 75.8 | 44 | 92.3 | 79.6 | 64.8 | 65.3 |
|   |   | Std | 4.9 | 4.7 | 2.7 | 2.9 | 5.5 | 4.3 | 4.1 | 3.2 |
|   | affa | Mean | 99.5 | 68.6 | 76.8 | 42.8 | 94.2 | 80.4 | 63.1 | 63.8 |
|   |   | Std | 4.7 | 4.3 | 4.1 | 3.8 | 4.5 | 2.4 | 6 | 3.8 |
|   | ava | Mean | 100.3 | 67.4 | 79.7 | 50.6 | 92.1 | 83.8 | 71.2 | 74.2 |
|   |   | Std | 4 | 4.2 | 4.8 | 5 | 4.6 | 3.3 | 6.4 | 4.6 |
|   | avva | Mean | 100.5 | 69.6 | 82.4 | 49.2 | 94.6 | 85.5 | 71.1 | 73.7 |
|   |   | Std | 3.9 | 3.4 | 4.7 | 5 | 3.4 | 3.9 | 5.9 | 4.4 |
|   | asa | Mean | 98.8 | 66.1 | 76.3 | 44.7 | 91.4 | 77.1 | 65.4 | 66.6 |
|   |   | Std | 3.9 | 3.8 | 2.5 | 2.8 | 4.7 | 2.5 | 4 | 2.2 |
|   | assa | Mean | 98.7 | 67.7 | 77.6 | 43.8 | 93.3 | 78.1 | 66.2 | 64.8 |
|   |   | Std | 4.4 | 4.1 | 3 | 2.8 | 4.1 | 3.2 | 3.1 | 3.1 |
| i | ifi | Mean | 91.3 | 59.1 | 78.6 | 46.8 | 83.9 | 76.8 | 69.6 | 67.4 |
|   |   | Std | 4 | 3.5 | 3 | 3 | 4 | 2.9 | 4.8 | 3.8 |
|   | iffi | Mean | 90.3 | 59.8 | 78.6 | 44.6 | 85.4 | 76.8 | 65.7 | 66.1 |
|   |   | Std | 4.3 | 3.5 | 4.8 | 4.5 | 3.9 | 2.2 | 7.1 | 5 |
|   | ivi | Mean | 90.9 | 58.4 | 81.4 | 52.1 | 82.3 | 80.7 | 73.3 | 74.2 |
|   |   | Std | 4.7 | 4.5 | 5.6 | 6 | 4.4 | 6.4 | 6.3 | 5.8 |
|   | ivvi | Mean | 90.2 | 59.4 | 82.7 | 49.7 | 84.2 | 80.9 | 71.3 | 72.9 |
|   |   | Std | 4.1 | 3.6 | 4.8 | 5.3 | 3.8 | 4.6 | 5.7 | 4.9 |
|   | isi | Mean | 89.7 | 57.4 | 76.1 | 43.9 | 81.3 | 75.2 | 67 | 65.5 |
|   |   | Std | 4.2 | 4.4 | 2.1 | 2.4 | 4.5 | 3.1 | 3.9 | 2.6 |
|   | issi | Mean | 88 | 57.1 | 77.6 | 43.5 | 82.1 | 75.6 | 66.3 | 65.5 |
|   |   | Std | 3.7 | 3.1 | 3.5 | 3.4 | 3.4 | 2.1 | 4.5 | 2.7 |
| u | ufu | Mean | 92.8 | 60.7 | 77.3 | 45.1 | 85.9 | 76.4 | 67.7 | 65.9 |
|   |   | Std | 3.6 | 2.8 | 2.8 | 3 | 3.4 | 3.7 | 3.8 | 2.8 |
|   | uffu | Mean | 93.1 | 62.4 | 77.4 | 43.4 | 87.8 | 78 | 64.4 | 65 |
|   |   | Std | 4.2 | 3.8 | 3.3 | 3.6 | 4.4 | 2.2 | 6 | 3.8 |
|   | uvu | Mean | 94.4 | 61.7 | 82.5 | 52.3 | 86.7 | 81.1 | 73.8 | 75.2 |
|   |   | Std | 3.6 | 3.6 | 5.9 | 6.4 | 4.1 | 6.2 | 7 | 7.1 |
|   | uvvu | Mean | 95.4 | 63.6 | 82.4 | 50.4 | 89.1 | 79.3 | 72.5 | 74.7 |
|   |   | Std | 3.4 | 3.1 | 4.9 | 5.2 | 3.4 | 4.1 | 5.7 | 5.2 |
|   | usu | Mean | 93.8 | 61.4 | 80.1 | 48.1 | 86.8 | 77.5 | 71.8 | 68.9 |
|   |   | Std | 4 | 3.9 | 2.9 | 2.5 | 4.6 | 3.6 | 3.3 | 4.3 |
|   | ussu | Mean | 91.3 | 60.3 | 81.5 | 47.4 | 85.7 | 76.6 | 71.8 | 67.9 |
|   |   | Std | 3.5 | 3.1 | 3.5 | 3.4 | 3.6 | 3.8 | 5.1 | 3.5 |

Table XXIX – Average and standard deviation of energy domain parameters for fricatives in singleton vs. geminate forms, averaged over speakers and repetitions (values in logarithmic form).

**Acknowledgments**

This work was funded by Sapienza University of Rome within research projects ("ex-quota 60%", "Ricerche di Facoltà") in the years 1991-2020 and supported in part by the Radcliffe Institute for Advanced Study at Harvard University. The authors wish to thank Maurizio Giovanardi and Simone Faluschi for their valuable contributions to the GEMMA project while they were interns in the Speech Lab at the DIET Department working toward their Master of Science degree, and Sara Budoni for her support in re-acquiring and converting recordings part of the section dedicated to fricatives in the GEMMA database.